\documentclass{ws-ijmpd}
\usepackage[super,compress]{cite}
\usepackage{color}
\usepackage{enumitem}
\usepackage{hyperref}
\usepackage{natbib}
\usepackage{aas_macros}
\usepackage{ragged2e}

\begin{document}

\markboth{M. Mezcua}
{Observational evidence for intermediate-mass black holes}

%
\catchline{}{}{}{}{}
%

\title{OBSERVATIONAL EVIDENCE FOR \\INTERMEDIATE-MASS BLACK HOLES}

\author{MAR MEZCUA}

\address{D\'epartement de Physique, Universit\'e de Montr\'eal, \\C.P. 6128, Succ. Centre-Ville, Montreal, Quebec H3C 3J7, Canada\\
Email: marmezcua.astro@gmail.com}

\maketitle


\begin{abstract}
Intermediate-mass black holes (IMBHs), with masses in the range $100-10^{6}$ M$_{\odot}$, are the link between stellar-mass BHs and supermassive BHs (SMBHs). They are thought to be the seeds from which SMBHs grow, which would explain the existence of quasars with BH masses of up to 10$^{10}$ M$_{\odot}$ when the Universe was only 0.8 Gyr old. The detection and study of IMBHs has thus strong implications for understanding how SMBHs form and grow, which is ultimately linked to galaxy formation and growth, as well as for studies of the universality of BH accretion or the epoch of reionisation. Proving the existence of seed BHs in the early Universe is not yet feasible with the current instrumentation; however, those seeds that did not grow into SMBHs can be found as IMBHs in the nearby Universe. In this review I summarize the different scenarios proposed for the formation of IMBHs and gather all the observational evidence for the few hundreds of nearby IMBH candidates found in dwarf galaxies, globular clusters, and ultraluminous X-ray sources, as well as the possible discovery of a few seed BHs at high redshift. I discuss some of their properties, such as X-ray weakness and location in the BH mass scaling relations, and the possibility to discover IMBHs through high velocity clouds, tidal disruption events, gravitational waves, or accretion disks in active galactic nuclei. I finalize with the prospects for the detection of IMBHs with up-coming observatories.
\end{abstract}



\section{INTRODUCTION}	
Black holes (BHs) were found to be the solution to the Einstein field equations of general relativity in the early 1960s\footnote{The metric describing a non-rotating BH with no charge (Schwarzschild BH) was found by Karl Schwarzschild in 1916, but it wasn't until 1963 and 1965 that Roy Kerr and Ezra Newman found the solutions for a rotating and a rotating plus electrically charged BH, respectively.} and, according to the no-hair theorem, they can be described by three parameters: mass, spin and charge. Independently of spin and charge, BHs are commonly classified into three types according to their mass:\\
\indent{} \textit{Stellar-mass BHs (3 M$_{\odot} < M_\mathrm{BH} \leq 100$ M$_{\odot}$)}. They are the end-product of a massive star ($>$15 M$_{\odot}$) that collapses into a BH when the star's fuel supply is burned out and the internal pressure is insufficient to support the gravitational force. The first solid evidence for the existence of BHs came from X-ray and optical observations in the 1970s and 1980s of X-ray binaries (XRBs; e.g., Cygnus X-1, LMC-X3; see reviews by \citealt{2006msu..conf..145C}; \citealt{2006ARA&A..44...49R}) whose compact object had a mass above 3 M$_{\odot}$ (i.e. too massive for a neutron star or a white dwarf; \citealt{1996A&A...305..871B}). Today, there are more than 20 confirmed stellar-mass BHs in XRBs.\\

\indent{} \textit{Supermassive BHs (SMBHs; $M_\mathrm{BH} \geq 10^{6}$ M$_{\odot}$)}. SMBHs are the most massive types of BHs and they reside at the center of most massive galaxies in the local Universe (see reviews by \citealt{2013ARA&A..51..511K}; \citealt{2013ApJ...764..184M}; \citealt{2016ASSL..418..263G}). The best observational evidence for a SMBH comes from studies of the proper motion of stars around the center of our Galaxy, which reveal the presence of a central BH with a mass of 4 $\times 10^{6}$ M$_{\odot}$ (\citealt{2008ApJ...689.1044G}; \citealt{2009ApJ...692.1075G}). The masses of SMBHs are observed to correlate with some of their host galaxy properties, such as bulge stellar mass, luminosity, or stellar velocity dispersion, suggesting a co-evolution or synchronized growth between galaxies and their central BHs (e.g. \citealt{1995ARA&A..33..581K}; \citealt{1998AJ....115.2285M}; \citealt{2000ApJ...539L...9F}; \citealt{2000ApJ...539L..13G}). SMBHs grow through the accretion of matter, during which they are observable as active galactic nuclei (AGN) or quasars. Quasars with BH masses of up to 10$^{10}$ M$_{\odot}$ have been detected when the Universe was only 0.8 Gyr old ($z \sim$ 7, e.g., \citealt{2001AJ....122.2833F,2003AJ....125.1649F}; \citealt{2007AJ....134.2435W,2010AJ....139..906W}; \citealt{2011Natur.474..616M}; \citealt{2013ApJ...779...24V}; \citealt{2015Natur.518..512W}). To reach this mass in such a short time, SMBHs should have started as lower-mass seed BHs of more than 100 M$_{\odot}$ at $z >$ 10 and grow very fast via accretion and mergers (e.g., \citealt{2010A&ARv..18..279V}; see Sect.~\ref{formation}).\\

\indent{} \textit{Intermediate-mass BHs (IMBHs; 100 M$_{\odot} < M_\mathrm{BH} < 10^{6}$ M$_{\odot}$)}. They are the link, thought to be missing for many decades, between stellar-mass and SMBHs and the possible seeds from which SMBHs in the early Universe grew. Finding proof of their existence is thus pivotal for understanding SMBH and galaxy growth. The study of their accretion physics and radiative properties is important for understanding the effects of BH feedback in the formation of the first galaxies and the quenching of star formation (e.g., \citealt{2011ApJ...738...54K}; \citealt{2012MNRAS.420.2662D}; \citealt{2012ApJ...754...34J}), for studies of the epoch of reionisation (\citealt{2004MNRAS.350..539R}), and for confirming whether accretion is a scale-invariant physical mechanism governing BHs of all masses. The latter has been already inferred from the fundamental plane of accreting BHs (e.g., \citealt{1995A&A...293..665F}; \citealt{2003MNRAS.345.1057M}; \citealt{2004A&A...414..895F}), which is a correlation between BH mass, X-ray luminosity (proxy of accretion flow) and radio luminosity (proxy of jet ejection) that extends all the way from stellar-mass to SMBHs, proving that a disk-jet coupling mechanism takes places in BHs of all masses (\citealt{2014ApJ...788L..22G}). Finally, IMBHs offer the best testbed for investigating tidal disruptions of stars (e.g., \citealt{2009MNRAS.400.2070S};  \citealt{2011ApJ...738L..13M}; see Sect.~\ref{TDEs}) and the coalescence of IMBHs pairs provide the right signal for detection in gravitational wave experiments (e.g., \citealt{2002MNRAS.331..805H}; \citealt{2010ApJ...722.1197A}; see Sect.~\ref{gravitationalwaves}). 

Proving the existence of primordial IMBHs at $z >$ 7 is a timely endeavour with the current facilities (e.g., \citealt{2015ApJ...808..139S}; \citealt{2016MNRAS.460.4003A}; \citealt{2016MNRAS.459.1432P}; \citealt{2016MNRAS.460.3143S}; \citealt{2017MNRAS.466.2131P}; \citealt{2017ApJ...838..117N}). Most of the studies of high-redshift BHs are limited to luminous quasars hosting SMBHs (see review by \citealt{2016PASA...33...54R}). However, observational evidence of those seed BHs that did not grow into SMBHs (the 'leftovers' of the early Universe) should be found in the local Universe (e.g., see review by \citealt{2012NatCo...3E1304G}; \citealt{2016PASA...33...54R}) and up to $z \sim$2.4 (\citealt{2016ApJ...817...20M}; Mezcua et al. in preparation): in dwarf galaxies, as because of their low mass and metallicity they resemble those galaxies formed in the early Universe (e.g., \citealt{2011Natur.470...66R}, \citeyear{2013ApJ...775..116R}, \citeyear{2014ApJ...787L..30R}; \citealt{2015ApJ...809L..14B,2017ApJ...836...20B}; \citealt{2016ApJ...817...20M}); in nearby globular clusters, as stellar clusters are one of the sites of possible IMBH formation (e.g., \citealt{2002ApJ...576..899P}; \citealt{2004Natur.428..724P}; \citealt{2006MNRAS.368..141F}; \citealt{2017Natur.542..203K}); or in the form of off-nuclear ultraluminous X-ray sources (ULXs) in the halos and spiral arms of large galaxies, as ULXs could be the stripped nucleus of dwarf galaxies (e.g., \citealt{2001ApJ...551L..27M}; \citealt{2009Natur.460...73F}; \citealt{2010ApJ...721L.148B}; \citealt{2013MNRAS.436.1546M}, \citeyear{2013MNRAS.436.3128M}, \citeyear{2015MNRAS.448.1893M}; \citealt{2014Natur.513...74P}). 

\begin{figure*}
 \includegraphics[width=\textwidth]{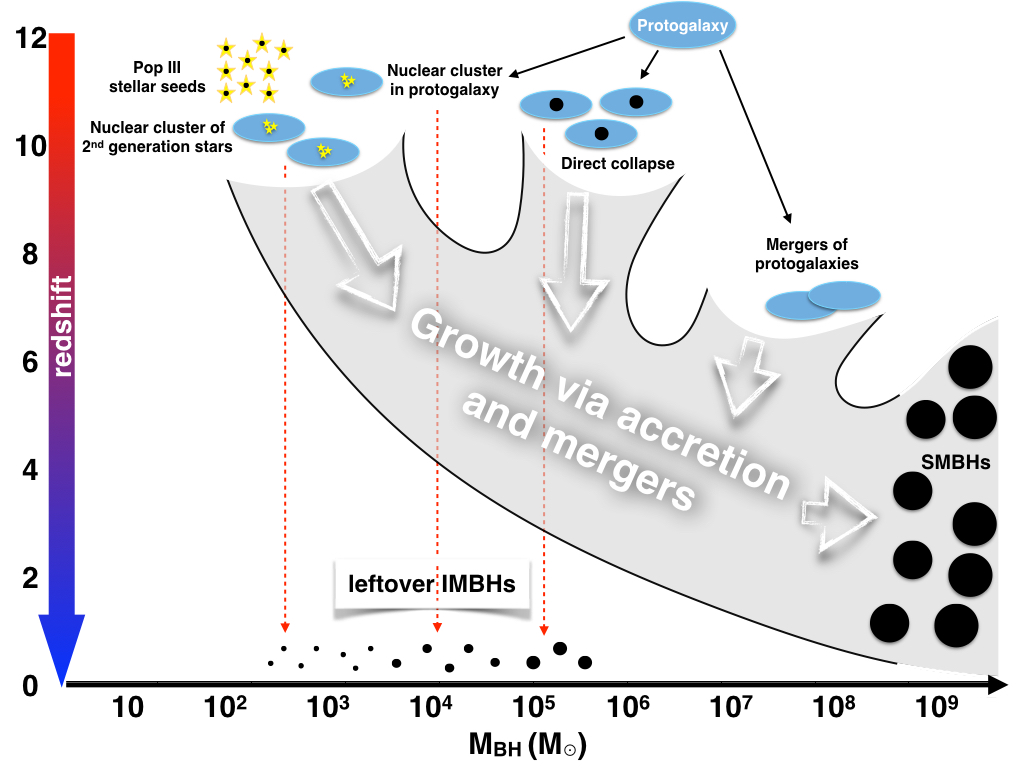}
 \caption{Formation scenarios for IMBHs. Seed BHs in the early Universe could form from Population III stars, from mergers in dense stellar clusters formed out either from the second generation of stars or from inflows in protogalaxies, or from direct collapse of dense gas in protogalaxies, and grow via accretion and merging to 10$^{9}$ M$_{\odot}$ by $z \sim$7. SMBHs could also directly form by mergers of protogalaxies at $z \sim$6. Those seed BHs that did not grow into SMBHs can be found in the local Universe as leftover IMBHs.}
 \label{formationmodels}
\end{figure*}

The aim of this review is to assemble all the observational evidence found so far for IMBHs. I will focus on globular clusters (Sect.~2.1), ULXs (Sect.~2.2), dwarf galaxies (Sect.~\ref{dwarf}), and seed BH candidates at $z > 6$ (Sect.~\ref{highz}), and mention other pathways for IMBH detection such as tidal disruption events, gravitational waves, accretion disks in AGN, and high velocity clouds (Sect.~\ref{others}). I will first start by providing a brief summary on seed BH formation (see Fig.~\ref{formationmodels}; for more details see reviews by e.g., \citealt{2010A&ARv..18..279V,2012Sci...337..544V}; \citealt{2012RPPh...75l4901V}; \citealt{2014GReGr..46.1702N}; \citealt{2016PASA...33....7J}; \citealt{2016PASA...33...51L}). 

Throughout the review I will use the terms 'seed BH' and 'IMBH' to refer to those BHs in the same mass regime (100 M$_{\odot} < M_\mathrm{BH} < 10^{6}$ M$_{\odot}$). The difference between the two relies on the redshift of the sources: 'seed BH' refers to the early Universe, while the term 'IMBH' will be used for lower redshift objects. Those BHs with $M_\mathrm{BH} \sim10^{5}-10^{6}$ M$_{\odot}$ are sometimes referred to as 'low-mass BHs' or 'low-mass AGN' in the literature (e.g., \citealt{2007ApJ...670...92G}; \citealt{2016ApJ...825..139P}), and other times as 'IMBHs' (e.g., \citealt{2007ApJ...656...84G}; \citealt{2012ApJ...761...73D}). Because of this disparity, in this review I define IMBHs as having 100 M$_{\odot} < M_\mathrm{BH} < 10^{6}$ M$_{\odot}$ and I use the term 'low-mass AGN' or 'low-mass BH' to refer to those BHs with $\sim10^{6}$ M$_{\odot}$.


\subsection{IMBHs: formation scenarios}
\label{formation}
BHs are fed by the accretion of gas in a process in which a small fraction of the energy of the accreted gas is released in the form of radiation. The luminosity at which the outward radiation balances the inward gravitational force is referred to as the Eddington luminosity, or Eddington limit, and can be written as:
\begin{equation}
L_\mathrm{Edd} = \frac{4\pi cGm_\mathrm{p}M_\mathrm{BH}}{\sigma_\mathrm{T}} \simeq 1.3 \times 10^{38} \left(\frac{M_\mathrm{BH}}{M_{\odot}}\right) \mathrm{erg/s}
\end{equation}
where $c$ is the speed of light, $G$ the gravitational constant, $m_\mathrm{p}$ the proton mass and $\sigma_\mathrm{T}$ the Thompson scattering cross-section. The Eddington rate is the rate at which a BH radiating at the Eddington luminosity is accreting mass from its surrounding. 

A seed BH of 100 M$_{\odot} < M_\mathrm{BH} < 10^{6}$ M$_{\odot}$ accreting at the Eddington rate would need more than 0.5 Gyr to reach $10^{9}$ M$_{\odot}$ (assuming a typical radiative efficiency of 10\%; \citealt{2010A&ARv..18..279V}). Therefore, the existence of SMBHs of more than 10$^{9}$ M$_{\odot}$ when the Universe was $\sim$1 Gyr old implies that the seed IMBHs formed at $z \geq 10$, in a primordial cold dark matter Universe in which dark matter halos grow out of the gravitational collapse of small density fluctuations. The first stars formed from the collapse of pristine (metal-free) gas in these dark matter halos. In the absence of metals (elements heavier than He and Li), gas cooling is only possible by means of atomic and molecular hydrogen (H$_{2}$; e.g., \citealt{1999ApJ...527L...5B,2002ApJ...564...23B}; \citealt{2004ARA&A..42...79B}). In such a early Universe, seed BHs could form from:\\

\indent{} \textit{(i) Population III stars.} If H$_{2}$ dominates the cooling rate, the primordial gas can cool down to $\sim$100 K and collapse into protostars (known as Pop III stars) of typically a few hundreds of solar masses (e.g., \citealt{2004ARA&A..42...79B}) and up to $\sim10^{3}$ M$_{\odot}$ (e.g., \citealt{2014ApJ...781...60H}). Only those Pop III stars over 260 M$_{\odot}$ collapse into a BH containing at least half of the initial stellar mass (i.e. $M_\mathrm{BH} \geq$100 M$_{\odot}$; \citealt{1984ApJ...280..825B}; \citealt{2001ApJ...550..372F}; \citealt{2002ApJ...567..532H}) and thus into IMBHs. However, the existence of numerous isolated stars more massive than 260 M$_{\odot}$ has been put into doubt by simulations showing that Pop III stars may instead form in binaries or multiple systems of 10--100 M$_{\odot}$ (e.g., \citealt{2009Sci...325..601T}; \citealt{2011Sci...331.1040C}; \citealt{2012MNRAS.422..290S}; see review by \citealt{2015ComAC...2....3G}). To reach a BH mass of 10$^{9}$ M$_{\odot}$ in $\sim$0.5 Gyr (the time elapsed between $z$ = 10 and $z$ = 6), Pop III stellar BH seeds would have to grow via supra-exponential accretion\footnote{Supra-exponential growth can describe Hoyle-Lyttleton wind accretion or spherical Bondi accretion (see e.g., \citealt{2017arXiv170100415A} and references therein for further details).} (e.g., when bound in a nuclear stellar cluster fed by flows of dense cold gas; \citealt{2014Sci...345.1330A}) or undergo phases of accretion at super-Eddington rates (e.g., \citealt{2005ApJ...633..624V}; \citealt{2014ApJ...784L..38M}; \citealt{2015MNRAS.451.1964S}; \citealt{2016MNRAS.458.3047P}). 

\indent{} \textit{(ii) Direct collapse.} IMBHs could also form inside the first metal-free (or very metal-poor) protogalaxies by direct collapse of rapidly inflowing dense gas (e.g., \citealt{1994ApJ...432...52L}; \citealt{1995ApJ...443...11E}; \citealt{2003ApJ...596...34B}; \citealt{2006MNRAS.371.1813L}). For the gas to reach the halo center and collapse to occur, fragmentation leading to star formation must be inhibited and the gas must have low angular momentum so that it undergoes gravitational instabilities instead of forming a rotationally-supported disk. The gravitational instabilities and inward gas transport can be achieved by the formation of bars within bars (\citealt{2006MNRAS.370..289B}), while fragmentation can be prevented if H$_{2}$ is destroyed by an intense Lyman-Werner (ultraviolet --UV) radiation and atomic hydrogen dominates the cooling rate. In such halos, gas cools down gradually only to $\sim10^{4}$ K and can form a supermassive star of $\sim10^{5}$ M$_{\odot}$. The collapse of such a supermassive star forms a BH of $\sim10^{4}-10^{5}$ M$_{\odot}$ (e.g., \citealt{2014MNRAS.443.2410F}), orders of magnitude more massive than Pop III stellar BH seeds, that can grow into a SMBH by $z \sim$7 without having to invoke super-critical accretion (e.g., \citealt{2016MNRAS.457.3356V}).
Given that the cosmic UV background may not be intense enough to prevent H$_{2}$ formation and gas fragmentation (\citealt{2008MNRAS.388...26J}), direct collapse can only occur in halos close (within 15 kpc) to luminous star-forming galaxies producing sufficient Lyman-Werner radiation (e.g., \citealt{2008MNRAS.391.1961D,2014MNRAS.442.2036D}; \citealt{2012MNRAS.425.2854A,2014MNRAS.443..648A}; \citealt{2016MNRAS.463..529H}) and is thus thought to be a much less common mechanism of IMBH formation. Previously-formed direct collapse seed BHs could also provide the radiation necessary to prevent star formation and form additional BHs, in which case they would be more abundant than expected if including only star-forming galaxies as the source of Lyman-Werner radiation (\citealt{2017ApJ...838..111Y}).

\indent{} \textit{(iii) Mergers in dense stellar clusters.}
Another way to form IMBHs is via runaway collisions of stars in dense stellar clusters (e.g., \citealt{1999A&A...348..117P,2004Natur.428..724P}; \citealt{2009ApJ...694..302D}; \citealt{2016MNRAS.459.3432M}). Compact nuclear stellar cluster can form out of the second generation of low-mass stars that formed from the gas metal-polluted (but with still highly sub-solar metallicity) by the first generation of Pop III stars (\citealt{2008ApJ...686..801O}). Frequent stellar mergers within the cluster can lead to the formation of a supermassive star that will collapse into a BH of $\sim10^{2}-10^{4}$ M$_{\odot}$ (\citealt{2009ApJ...694..302D}). Dense stellar clusters might also form at the center of the protogalaxies previously described when the central density is increased by the inflow of the metal-poor gas. The mass of these clusters is typically of $\sim10^{5}$ M$_{\odot}$ and runaway stellar collisions can yield the formation of a supermassive star that will collapse into a BH of $\sim10^{3}$ M$_{\odot}$ (\citealt{2010MNRAS.409.1057D,2012MNRAS.421.1465D}).

\indent{} \textit{(iv) Other models: direct SMBH formation.}
Instead of seed BHs at $z > 10$ having to grow through accretion and mergers to 10$^{9}$ M$_{\odot}$ by $z \sim$7, the existence of SMBHs in such a young Universe can be also explained if these formed directly by mergers of massive protogalaxies at $z \sim$5--6 (\citealt{2010Natur.466.1082M}, \citeyear{2015ApJ...810...51M}; \citealt{2014MNRAS.437.1576B}; but see also \citealt{2013MNRAS.434.2600F}). When the two protogalaxies merge, merger-driven inflows of metal-enriched gas produce a massive ($\geq 10^{9}$ M$_{\odot}$) compact nuclear gas disk with a high angular momentum. In the inner parsecs an ultra-dense massive disky core is formed, which can turn into a supermassive star and collapse directly into a SMBH of 10$^{8}$--10$^{9}$ M$_{\odot}$ (\citealt{2015ApJ...810...51M}). Although this avenue of SMBH formation requires initial conditions more complex than those of the direct collapse in less massive halos scenario, it offers an explanation for the existence of high-z SMBHs without having to prevent gas cooling and star formation nor requiring primordial gas composition. 

\begin{figure*}
 \includegraphics[width=0.7\textwidth]{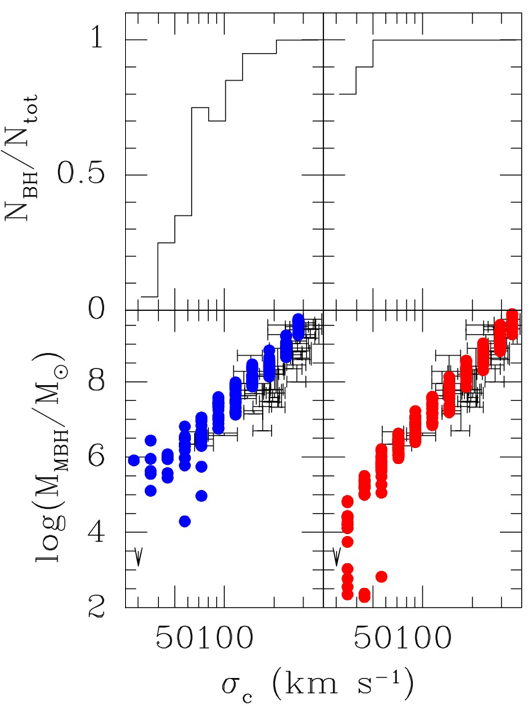}
 \caption{ \textbf{Top:} Predicted fraction of galaxies that host a central BH at a given velocity dispersion for the direct collapse (left) and the Pop III (right) seed BH formation models. \textbf{Bottom:} Predicted $M_\mathrm{BH}-\sigma$ relation at $z = 0$ starting from heavy direct collapse seed BHs (left panel, blue dots) and light Pop III seed BHs (right panel, red dots). Observational data (from \citealt{2002ApJ...574..740T}) are shown by their quoted error bars both in $\sigma$ and $M_\mathrm{BH}$. Figure and caption adapted from \cite{2010A&ARv..18..279V}. Reproduced with permission. \copyright IAU.}
 \label{occupationfraction}
\end{figure*}

Since the detection of $z > $7 seed BHs is yet challenging with current instrumentation, determining which (if any) of the above seed BH formation scenarios is correct requires to construct models of high-z seed BH formation, predict the leftover populations of IMBHs at low redshift for each scenario, and compare these to the observed number of IMBHs so far available. In the local Universe, leftover IMBHs are expected to reside in dwarf star-forming galaxies, as these have undergone a quieter merging/accretion history than massive galaxies and are thus more likely to resemble the primordial low-metallicity galaxies of the infant Universe. Given that in the early Universe Pop III seed BHs were presumably much more abundant than direct collapse seed BHs, simulations predict a higher BH occupation fraction in today's dwarf galaxies if the Pop III scenario was the dominant seeding mechanism at $z > $10 (\citealt{2008MNRAS.383.1079V,2010A&ARv..18..279V}; \citealt{2010MNRAS.408.1139V}; \citealt{2012NatCo...3E1304G}; see Fig.~\ref{occupationfraction}, top). The different seed BH formation scenarios should also leave an imprint on the tight correlations found between SMBH mass and host galaxy properties: light Pop III seeds are predicted to be undermassive with respect to the $M_\mathrm{BH}-\sigma$ relation, while the more massive direct collapse seed BHs are expected to lie above it (\citealt{2008MNRAS.383.1079V}, \citealt{2010A&ARv..18..279V}; see Fig.~\ref{occupationfraction}, bottom). At the high-mass end, no differences are predicted by the different seeding scenarios -- in these systems the initial conditions (seed BH mass) have been erased as a consequence of several mergers and accretion phases. Observationally, the $M_\mathrm{BH}-\sigma$ and $M_\mathrm{BH}-M_\mathrm{bulge}$ correlations are indeed found to be very strong for SMBHs of $10^{6}-10^{9}$ M$_{\odot}$, but they seem to bend or to have a large scatter both at the highest (e.g., \citealt{2011Natur.480..215M}; \citealt{2012MNRAS.424..224H}; \citealt{2015ApJ...808...79F}) and lowest (e.g., \citealt{2012NatCo...3E1304G}; \citealt{2013ApJ...764..151G,2015ApJ...798...54G}; \citealt{2015ApJ...813...82R}; \citealt{2016ApJ...817...21S}; see Sect.~\ref{dwarf} and Fig.~\ref{Msigma}) mass regimes (although see \citealt{2016ApJ...818..172G,2016ApJ...819...43G}; \citealt{2016MNRAS.457..320S}). Finding observational evidence of IMBHs in the local Universe, deriving their occupation fraction, and measuring the BH mass in the low-mass regime (i.e. in dwarf galaxies) is thus pivotal for understanding how seed BHs formed in the early Universe and evolved until the SMBHs observed today.

\section{OBSERVATIONAL EVIDENCE FOR IMBHs}	
To probe the existence of IMBHs we need to measure their BH mass. The use of stellar or gas dynamics is the most secure way to weight BHs; however, the sphere of influence of a BH of $10^{5}$ M$_{\odot}$ is of only 0.5 pc and cannot be resolved, with the current instrumentation, beyond $\sim$1 Mpc. Dynamical BH masses in the intermediate-mass regime have thus only been obtained for nearby dwarf galaxies (most within the Local Group) and in globular clusters. In the absence of kinematic signatures, radiative signatures of BH accretion (e.g., X-ray and radio emission) must be used to infer the presence of IMBHs and estimate their BH mass. These methods have provided the detection of IMBHs in globular clusters, ultraluminous X-ray sources, and dwarf galaxies in the local Universe and up to $z \sim$2. At higher redshifts, the near-infrared (NIR) detection of a strong Ly$\alpha$ emission line and the combination of NIR photometry with deep X-ray observations has yielded the identification of a few direct collapse BH candidates.

\subsection{Globular clusters}
\label{globular}
One of the possible formation scenarios for IMBHs is the runaway core collapse and coalescence of stars in stellar clusters. Globular clusters have thus been common targets in the search for IMBHs. The presence of IMBHs in globular clusters was first suggested by \cite{1975ApJ...200L.131S} when studying their X-ray emission. They argued that the flux of the globular cluster X-ray sources could be explained by accretion onto a 100-1000 M$_{\odot}$ BH, which was supported by the finding of high central escape velocities. Since then many studies have aimed at detecting IMBHs in globular clusters through their radiative accretion signatures, but no conclusive results have been obtained (e.g., \citealt{2005MNRAS.356L..17M}; \citealt{2008AJ....135..182B}; \citealt{2008MNRAS.389..379M}; \citealt{2010MNRAS.406.1049C}; \citealt{2012ApJ...750L..27S}; \citealt{2013ApJ...773L..31H}; \citealt{2015AJ....150..120W}). The only globular cluster with detected X-ray and radio emission is G1, in M31 (\citealt{2006ApJ...644L..45P}; \citealt{2007ApJ...661..875K}; \citealt{2007ApJ...661L.151U}), suggesting the presence of an IMBH with a BH mass of $(1.8 \pm 0.5) \times 10^{4}$ M$_{\odot}$ estimated from photometric and kinematic observations (\citealt{2002ApJ...578L..41G,2005ApJ...634.1093G}). However, later results obtained by \cite{2012ApJ...755L...1M} did not detect any radio emission. Globular clusters have little gas and dust and thus any signatures of accretion from a putative IMBH in the X-ray and radio regimes are expected to be very low. This could explain the lack of X-ray and radio detections, leaving the kinematic signatures as currently the most viable method of probing the presence of IMBHs in these stellar systems.  

\begin{table}[t]
\tbl{IMBH candidates in globular clusters}
{\begin{tabular}{@{}lccc@{}} \toprule
Name		 		& 				&	M$_\mathrm{BH}$			&	References  							\\
					&				&        [M$_{\odot}$]  			&  										\\ \colrule
47 Tuc				&				&   $2.2^{+1.5}_{-0.8} \times 10^{3}$	&  	\cite{2017Natur.542..203K}						\\
G1$^{\dagger}$			&				&	$(1.8 \pm 0.5) \times 10^{4}$	& 	\cite{2005ApJ...634.1093G}					\\
NGC 1851			&				&	$< 2 \times 10^{3}$			&  	L\"utzgendorf et al. (2013)					\\
NGC 1904			&		M79		&      $(3 \pm 1) \times 10^{3}$		&	L\"utzgendorf et al. (2013)						\\
NGC 2808			&				&      $1 \times 10^{4}$			&	L\"utzgendorf et al. (2012)						\\
NGC 5139			&	$\omega$Cen  &	$(4.7 \pm 1.0) \times 10^{4}$    &	\cite{2010ApJ...719L..60N}						\\
NGC 5286			&				&	$(1.5 \pm 1.0) \times 10^{3}$	&	Feldmeier et al. (2013)					\\
NGC 5694			&				&	$< 8 \times 10^{3}$			&	L\"utzgendorf et al. (2013)						\\
NGC 5824			&				& 	$< 6 \times 10^{3}$			&	L\"utzgendorf et al. (2013)						\\ 
NGC 6093			&		M80		&	$< 8 \times 10^{2}$			&	L\"utzgendorf et al. (2013)						\\ 
NGC 6266			&		M62		&	$(2 \pm 1) \times 10^{3}$		&	L\"utzgendorf et al. (2013)						\\ 
NGC 6388			&				&	$(2.8 \pm 0.4) \times 10^{4}$	&	L\"utzgendorf et al. (2015)						\\
NGC 6715			&		M54		&	$9.4 \times 10^{3}$			&	\cite{2009ApJ...699L.169I}						\\ 
NGC 7078			&		M15		&	$(3.9 \pm 2.2) \times 10^{3}$	&	\cite{2002AJ....124.3270G}					\\ \botrule
\end{tabular} \label{tableglobular}}
$^{\dagger}$ This is the only globular cluster with X-ray and radio signatures of BH accretion (\citealt{2006ApJ...644L..45P}; \citealt{2007ApJ...661..875K}; \citealt{2007ApJ...661L.151U}; but see \citealt{2012ApJ...755L...1M}).
\end{table}

The first IMBH candidate in a globular cluster based on kinematic measurements and dynamical modeling was M15 (\citealt{1976ApJ...209..214B}; \citealt{1989ApJ...347..251P}), which presented a pronounced rise in its velocity dispersion profile and for which a BH mass of $(3.9 \pm 2.2) \times 10^{3}$ M$_{\odot}$ was estimated (\citealt{1997AJ....113.1026G}, \citeyear{2000AJ....119.1268G}; \citealt{2002AJ....124.3270G}). However, the results could also be explained by a central concentration of compact objects (e.g., \citealt{2003ApJ...582L..21B}; \citealt{2006ApJ...641..852V}), which weakened the IMBH scenario for this globular cluster. The use of integral field spectroscopy, to obtain the central velocity-dispersion profile, and of photometric data (e.g., with the \textit{Hubble Space Telescope}, \textit{HST}), to obtain the cluster photometric center and surface brightness profile, has allowed estimating the BH mass in a dozen more globular clusters by comparing the data to spherical dynamical models. This is the case of another strong IMBH candidate in a globular cluster, $\omega$ Centauri, for which \cite{2008ApJ...676.1008N,2010ApJ...719L..60N} claimed the presence of an IMBH of best-fitted mass $(4.7 \pm 1.0) \times 10^{4}$ M$_{\odot}$ while \cite{2010ApJ...710.1063V} reported an upper limit of $1.2 \times 10^{4}$ M$_{\odot}$. \cite{2017MNRAS.464.2174B} also found that the velocity dispersion profile of $\omega$ Centauri is best fitted by an IMBH of $10^{4}$ M$_{\odot}$. Using integral-field spectroscopy and \textit{HST} photometry, \citeauthor{2012A&A...542A.129L} (2012, \citeyear{2013A&A...552A..49L}, \citeyear{2015A&A...581A...1L}) reported upper limits on the mass of a putative BH in the globular clusters NGC 1851, NGC 2808, NGC 5694, NGC 5824, and NGC 6093 (see Table~\ref{tableglobular}) and predicted the presence of an IMBH of $(3 \pm 1) \times 10^{3}$ M$_{\odot}$ in NGC 1904, of $(2 \pm 1) \times 10^{3}$ M$_{\odot}$ in NGC 6266, and of $(2.8 \pm 0.4) \times 10^{4}$ M$_{\odot}$ in NGC 6388. An IMBH of $(1.5 \pm 1.0) \times 10^{3}$ M$_{\odot}$ is also suspected in the globular cluster NGC 5286 (\citealt{2013A&A...554A..63F}) and \cite{2009ApJ...699L.169I} reported the possible presence of an IMBH of $\sim$9400 M$_{\odot}$ in NGC 6715 (M54), a globular cluster located at the center of the Sagittarius dwarf galaxy. Nonetheless, no observational evidence for accretion in the form of X-ray or radio emission has been detected for any of these IMBH candidates, indicating that the globular clusters are devoid of gas within the BH radius of influence.

\cite{2017Natur.542..203K} recently proposed to use measurements of pulsar accelerations, which show an additional component beyond that caused by the gravitational potential of the cluster, together with N-body simulations to obtain stringent constraints on the central BH mass of globular clusters.  They apply this method to the globular cluster 47 Tuc, which hosts 25 known millisecond pulsars (\citealt{2001MNRAS.326..901F}; \citealt{2016MNRAS.459L..26P}; \citealt{2016MNRAS.462.2918R}), and find that those models with an IMBH produce pulsar accelerations more consistent with the observed accelerations than models without an IMBH. They infer a BH mass for the IMBH of $2.2 \times 10^{3}$ M$_{\odot}$ and provide an independent measure of the cluster mass ($0.75 \times 10^{6}$ M$_{\odot}$) that is in agreement with kinematic results (\citealt{2017MNRAS.464.2174B}). 
 This method is a promising way to infer the presence of an IMBH in those globular clusters whose lack of gas in their cores does not permit an electromagnetic detection of the BH. 
 Alternative methods include gravitational waves (e.g., \citealt{2010ApJ...719..987M}) and gravitational microlensing (e.g., \citealt{2016MNRAS.460.2025K}); however, these have not yet provided any IMBH candidates.

\subsection{Ultraluminous X-ray sources}
\label{ULXs}
IMBHs were also proposed to explain the nature of ultraluminous X-ray sources (ULXs). ULXs are extragalactic and off-nuclear X-ray sources with luminosities $L_\mathrm{X} \geq 10^{39}$ erg s$^{-1}$, which corresponds to the Eddington limit for a 10 M$_{\odot}$ stellar-mass BH. ULXs could thus host BHs of intermediate masses if accreting isotropically below the Eddington rate (e.g., \citealt{1999ApJ...519...89C}). Alternatively, they could be powered by stellar-mass BHs or magnetized neutron stars with near to super-Eddington accretion (e.g., see reviews by \citealt{2011NewAR..55..166F}; \citealt{2013MmSAI..84..629G}; \citealt{2015mbhe.confE..30B}; \citealt{2016AN....337..534R}; \citealt{2017arXiv170310728K}).  

The first evidence for the presence of IMBHs in ULXs came from the fit of the X-ray spectrum by a cool disk (soft excess of 0.1-0.3 keV) plus a power-law tail: if coming from a standard BH, the inverse proportionality between disk temperature and BH mass (\citealt{1973A&A....24..337S}) implied masses $\sim10^{4}$ M$_{\odot}$ (e.g., \citealt{2003ApJ...585L..37M}). However, the later finding of a curved spectrum with a cutoff above a few keV (e.g., \citealt{2006MNRAS.368..397S}; \citealt{2013ApJ...778..163B}; \citealt{2013ApJ...773L...9W,2014ApJ...793...21W}; \citealt{2014Natur.514..198M}; \citealt{2015ApJ...799..121R}) weakened the interpretation of a standard BH with an intermediate mass and supported the scenario in which most ULXs (those with X-ray luminosities $< 5 \times 10^{40}$ erg s$^{-1}$) are stellar-mass BHs in super-Eddington accretion regimes (known as the ultraluminous state; \citealt{2009MNRAS.397.1836G}). Dynamical evidence for the presence of a stellar-mass BH has been found in some ULXs (M101-X1, \citealt{2013Natur.503..500L}; NGC 7793 P13, \citealt{2014Natur.514..198M}, but later found to host a neutron star, \citealt{2016ApJ...831L..14F}, \citealt{2017MNRAS.466L..48I}), though the most surprising case is that of M82 X-2. M82 X-2 was thought to host an IMBH of more than 10$^{5}$ M$_{\odot}$ (\citealt{2010ApJ...710L.137F}) based on its high X-ray luminosity (peak at $3 \times 10^{40}$ erg s$^{-1}$), strong variability on scales of weeks, and low-frequency quasi-periodic oscillations (QPOs; \citealt{2007ApJ...671..349K}; \citealt{2007ApJ...668..941F}; \citealt{2010ApJ...710L.137F}). However, the finding of X-ray pulsations indicates that M82 X-2 is a neutron star (\citealt{2014Natur.514..202B}). This was the first discovery of a neutron star hosted by a ULX, an unexpected scenario for which there are two more known cases (\citealt{2016ApJ...831L..14F}; \citealt{2017Sci...355..817I,2017MNRAS.466L..48I}) and which could explain the nature of many more ULXs (\citealt{2017MNRAS.468L..59K}).

\begin{table}[t]
\tbl{Strong IMBH candidates among ULXs}
{\begin{tabular}{@{}lccc@{}} \toprule
Name		 		& Peak $L_\mathrm{X}$			&	M$_\mathrm{BH}$		&	Properties  												\\
					&	[erg s$^{-1}$]  				&        [M$_{\odot}$]  		&  															\\ \colrule
HLX-1				&	$1.1 \times 10^{42}$			&	$(0.3-30) \times 10^{4}$	& Outbursts, spectral state transitions, 								\\
					&							&						& jet radio emission, optical counterpart.	 							\\
M82-X1				&	$1.1 \times 10^{41}$			&	428 $\pm$ 105			& Variable, QPOs, spectral state transitions, 							\\
					& 							&						& optical counterpart.								 			 \\
NGC 2276-3c			& 6 $\times$ 10$^{40\dagger}$		&	5 $\times$ 10$^{4}$		& Variable, jet radio emission. 										 \\
CXO J122518.6+144545  &	$2.2 \times 10^{41}$			&		?				& Outbursts, optical counterpart.				
									\\ \botrule
\end{tabular} \label{tableULXs}}
$^{\dagger}$ When blended with two other ULXs. \\
References. HLX-1: \cite{2009Natur.460...73F}, 2012; \cite{2011ApJ...735...89L}; \cite{2011ApJ...743....6S}; \cite{2012Sci...337..554W}; \cite{2014ApJ...793..105G}. M82-X1: \cite{2001MNRAS.321L..29K}; \cite{2001ApJ...547L..25M}; \cite{2003ApJ...586L..61S}; \cite{2007ApJ...668..941F,2010ApJ...712L.169F}; \cite{2015ApJ...812L..34W}. NGC 2276-3c: \cite{2012MNRAS.423.1154S}; \citeauthor{2015MNRAS.448.1893M} (\citeyear{2013MNRAS.436.3128M}, 2015). CXO J122518.6+144545: \cite{2010MNRAS.407..645J}; \cite{2015MNRAS.454L..26H}.
\end{table}

\begin{figure*}
\includegraphics[width=\textwidth]{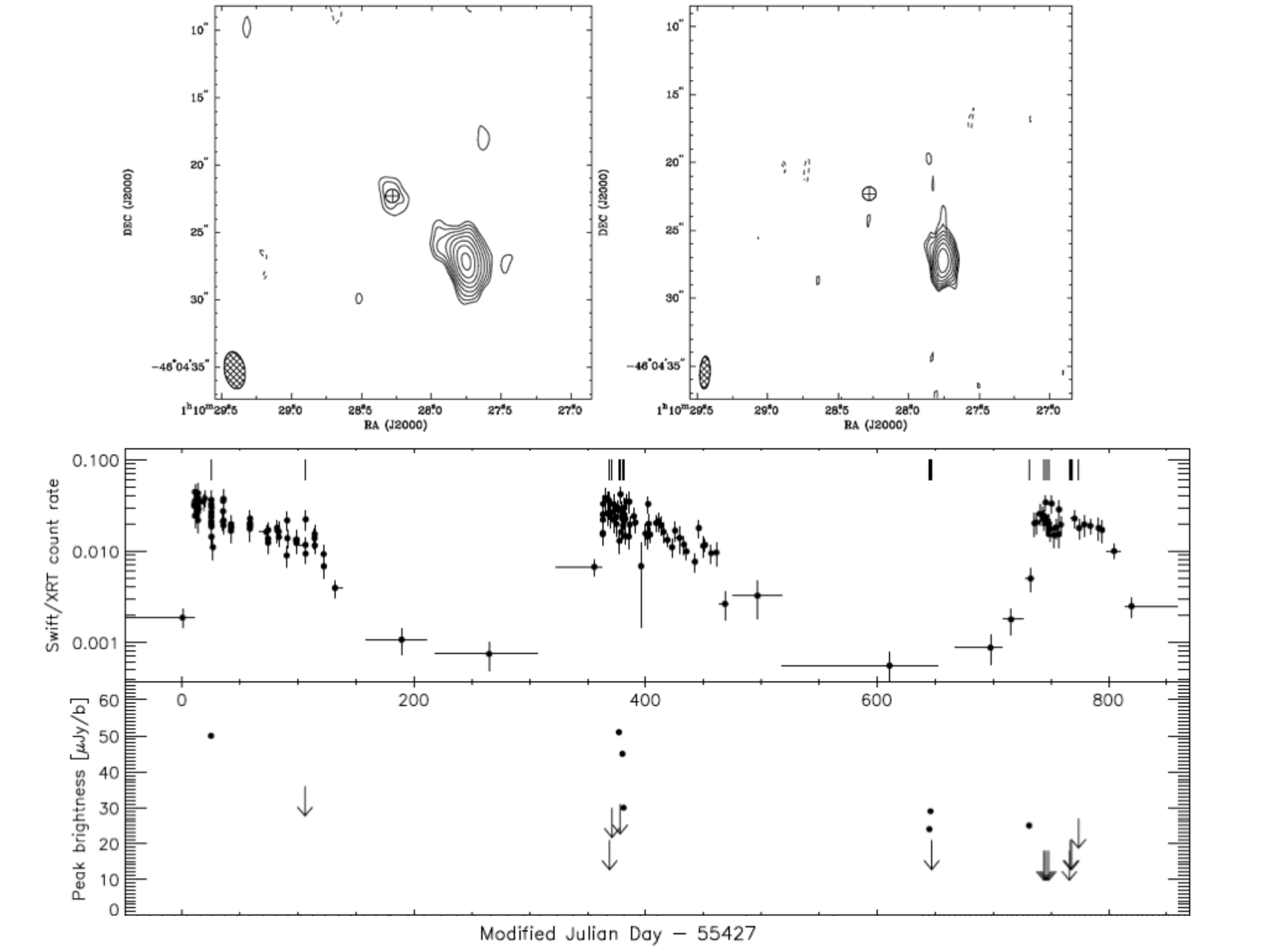}
\caption{\textbf{Top left}: Australia Telescope Compact Array (ATCA) image of HLX-1 and its host galaxy, ESO 243-49, during the X-ray hard state at 6.8 GHz central frequency. Radio contours start at $\pm$3 times the rms noise level of 3.3 $\mu$Jy beam$^{-1}$ and increase as 9.9 $\times (\sqrt{2})^n$ $\mu$Jy beam$^{-1}$. The Gaussian restoring beam size (shown in the lower left corner) is 2.79 arcsec $\times$ 1.53 arcsec at a major axis position angle of 10$^{\circ}$. \textbf{Top right}: Karl G. Jansky Very Large Array (VLA) image of the region of HLX-1 showing its host galaxy at 7.25 GHz central frequency. Radio contours start at $\pm$3 times the rms noise level of 2.1 $\mu$Jy beam$^{-1}$ and increase as 6.3 $\times (\sqrt{2})^n$ $\mu$Jy beam$^{-1}$. The Gaussian restoring beam size (shown in the lower left corner) is 2.40 arcsec $\times$ 0.76 arcsec at a major axis position angle of -3$^{\circ}$. The circles with a cross shows the 95\% positional uncertainty of the \textit{Chandra} and \textit{HST} counterpart of HLX-1 with a radius of 0.5 arcsec (\citealt{2012ApJ...747L..13F}). \textbf{Bottom}: Evolution of the 2010, 2011, and 2012 outburst of HLX-1 along with radio measurements. The top panel shows the \textit{Swift} X-ray lightcurve with count rates in the 0.3-10 keV band. The ticks above each outburst indicate the epochs of radio observations. The bottom panel shows the corresponding radio measurements. Detections are indicated with dots and non-detections (values of 3$\sigma$ upper limits using the rms noise level) with arrows. Figures and caption from \cite{2015MNRAS.446.3268C}. Reproduced with permission from the RAS.}
\label{HLX1}
\end{figure*}

Extreme ULXs, with $L_\mathrm{X} \geq 5 \times 10^{40}$ erg s$^{-1}$, and hyperluminous X-ray sources (HLXs), with $L_\mathrm{X} \geq 10^{41}$ erg s$^{-1}$, remain as the best candidates to IMBHs as their high X-ray luminosities can be difficult to explain even by super-Eddington accretion (see Table~\ref{tableULXs}). This is the case for HLX-1, the most well-known off-nuclear IMBH candidate. HLX-1 has an isotropic X-ray luminosity of $10^{42}$ erg s$^{-1}$ (\citealt{2009Natur.460...73F}) and, similarly to XRBs, it undergoes periodic outbursts over a timescale of months (\citealt{2011ApJ...735...89L}; \citealt{2011ApJ...743....6S}; \citealt{2014ApJ...793..105G}; \citealt{2015ApJ...811...23Y}) during which spectral state transitions over a timescale of days occur (\citealt{2009ApJ...705L.109G}; \citealt{2011ApJ...735...89L}; \citealt{2011ApJ...743....6S}; \citealt{2012Sci...337..554W}). The source also exhibits transient jet radio emission following the transition from a powerlaw-like spectrum (X-ray low/hard state) to a thermal-shape spectrum (X-ray high/soft state; \citealt{2012Sci...337..554W}; \citealt{2015MNRAS.446.3268C}; see Fig.~\ref{HLX1}). From the radio emission and applying the fundamental plane of accreting BHs, a BH mass of $9 \times (10^{3} - 10^{4})$ was estimated by \cite{2012Sci...337..554W} and an upper limit of $2.8 \times 10^{6}$ M$_{\odot}$ by \cite{2015MNRAS.446.3268C}. \cite{2015MNRAS.446.3268C} suggested that the radio emission is Doppler-boosted and that HLX-1 could be an outlier on the fundamental plane. Based on Eddington arguments, \cite{2011ApJ...743....6S} set a lower limit on the BH mass of 9 $\times10^{3}$ M$_{\odot}$, while using accretion disk models the BH mass was estimated to be in the range $6 \times 10^{3}$ M$_{\odot}$ $< M_\mathrm{BH} < 3 \times 10^{5}$ M$_{\odot}$ (\citealt{2011ApJ...734..111D}; \citealt{2012ApJ...752...34G}; \citealt{2014A&A...569A.116S}). The redshift of the optical counterpart confirms the association of HLX-1 with the host lenticular galaxy ESO 243-49 and suggests that it could be the nucleus of a dwarf galaxy that underwent a minor merger with ESO 243-49 (\citealt{2010ApJ...721L.102W}; \citealt{2012ApJ...747L..13F}; \citealt{2013ApJ...768L..22S}). 

\begin{figure*}
\includegraphics[width=\textwidth]{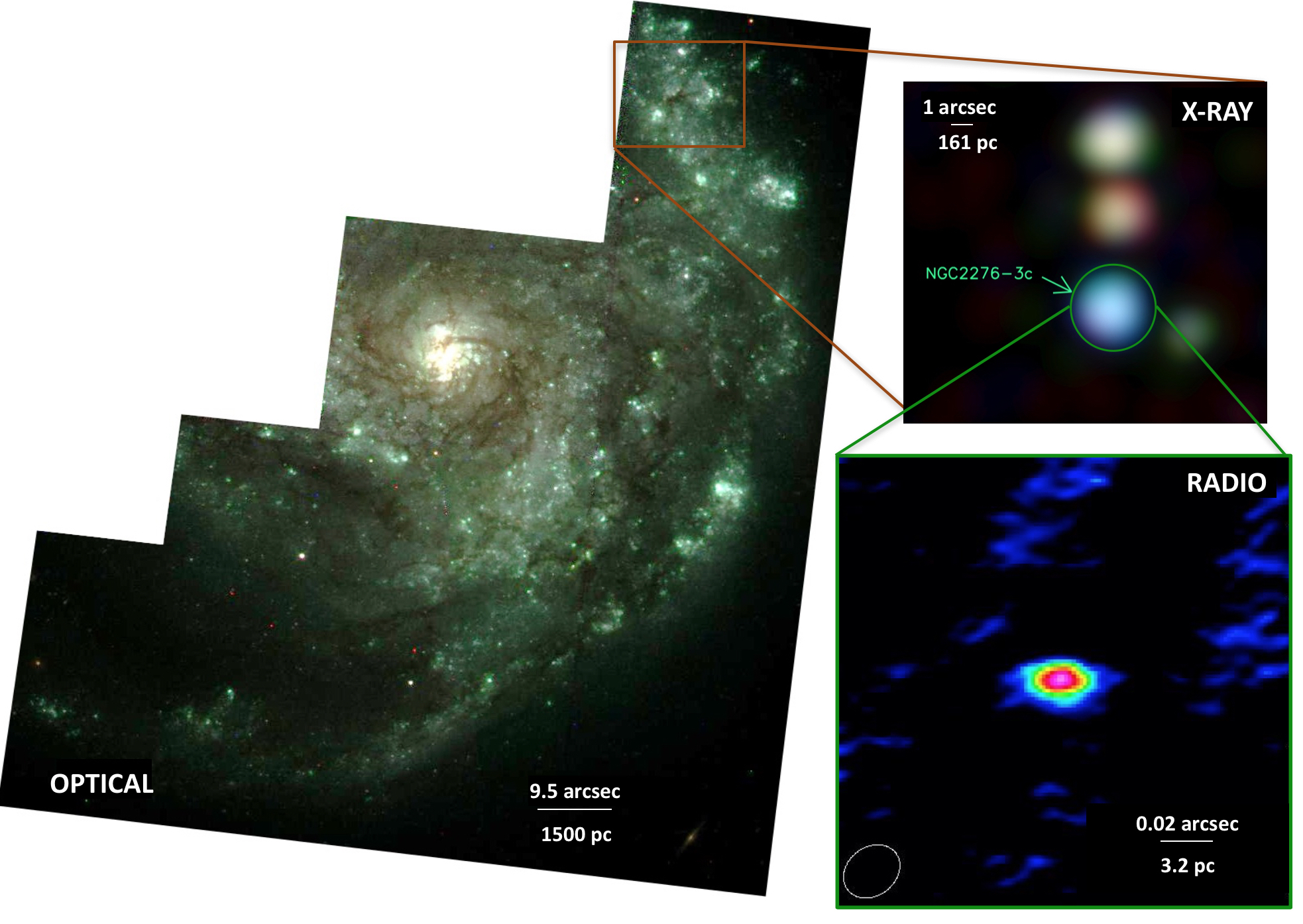}
\caption{\textbf{Optical}: Three-color image of the western arm of NGC\,2276 where the ULX NGC 2276-3c is located. The image has been created using three filters of the Wide Field Planetary Camera 2 (WFPC2) on the \textit{HST}: \textit{F814W} in red ($\sim$0.8 $\mu$m), \textit{F606W} in green ($\sim$0.6 $\mu$m) and \textit{F550W} in blue ($\sim$0.55 $\mu$m). \textbf{X-ray}: \textit{Chandra} X-ray satellite red (0.2--1.5 keV), green (1.5--2.5 keV), blue (2.5--8 keV) image convolved with a $\sim$2 arcsec full width at half maximum gaussian. The position of NGC 2276-3c is marked with a green circle. \textbf{Radio}: European VLBI Network 1.6 GHz image of NGC 2276-3c. The synthesized beam size is 16.4 mas $\times$ 13.1 mas oriented at a position angle of $-52^{\circ}$. The off-source rms noise is 8 $\mu$Jy beam$^{-1}$. The emission has a flux density of 65 $\pm$ $\mu$Jy. Image and caption adapted from \cite{2015MNRAS.448.1893M}. Reproduced with permission from the RAS.}
\label{NGC2276-3c}
\end{figure*}

M82 X-1 is the second strongest IMBH among HLXs because of its variability (Ptak \& Griffiths 1999; Kaaret \& Feng 2007), peak X-ray luminosity above $10^{41}$ erg s$^{-1}$ (\citealt{2001MNRAS.321L..29K}; \citealt{2001ApJ...547L..25M}), spectral transitions similar to those of standard BHs (\citealt{2010ApJ...712L.169F}), and low-frequency  QPOs (\citealt{2003ApJ...586L..61S}; \citealt{2007ApJ...668..941F}). Based on the finding of twin-peak QPOs at $\sim$3 Hz and 5 Hz and extrapolating the inverse scaling between BH mass and frequency that holds for stellar-mass BHs, the BH mass of M82 X-1 was estimated to be 428 $\pm$ 105 M$_{\odot}$ (\citealt{2014Natur.513...74P}). M81 X-1 was also suggested to be the nucleus of a stripped galaxy (\citealt{2005MNRAS.357..275K}), as is the case of the extreme ULX NGC 2276-3c, located in a peculiar arm of the spiral galaxy NGC 2276 (\citealt{2012MNRAS.423.1154S}; \citeauthor{2015MNRAS.448.1893M} \citeyear{2013MNRAS.436.3128M}, 2015). 
NGC 2276-3c was detected by the \textit{XMM-Newton} satellite as an extreme ULX of $L_\mathrm{X}\sim$6 $\times$ 10$^{40}$ erg s$^{-1}$ blended with two other ULXs (\citealt{2012MNRAS.423.1154S}). Later observations with the \textit{Chandra} X-ray satellite were able to resolve the source, which is strongly variable and whose spectral modeling was consistent with the sub-Eddington hard X-ray state (\citealt{2015MNRAS.448.1893M}). Very long baseline interferometry (VLBI) radio observations were performed quasi-simultaneously (one day difference) to the \textit{Chandra} observations, revealing a radio jet, characteristic of the hard state, with a size several orders of magnitude larger than the typical jet size of stellar-mass BHs but smaller than those of SMBH (Fig.~\ref{NGC2276-3c}; \citealt{2015MNRAS.448.1893M}; though Yang et al. 2017 failed to confirm the detection). A BH mass of 5 $\times$ 10$^{4}$ M$_{\odot}$ was estimated using the fundamental plane for accreting BHs, consistent with NGC 2276-3c containing an IMBH in the hard state (\citealt{2015MNRAS.448.1893M}).
The detection of compact radio jets has suggested the presence of IMBHs in some other ULXs for which the fundamental plane provides a BH mass estimate in the IMBH regime (e.g., N4861-X2, N4088-X1, \citealt{2011AN....332..379M}, \citealt{2014ApJ...785..121M}; IC342 X-1, \citealt{2012ApJ...749...17C}; N5457-X9, \citealt{2013MNRAS.436.1546M}). However, the luminosity of these ULXs being $< 5 \times 10^{40}$ erg s$^{-1}$ makes them also consistent with the super-Eddington accretion scenario. The presence of jets in the super-Eddington regime is expected from simulations (\citealt{2015MNRAS.454L...6M}) and was detected for the ULX Holmberg II X-1, which is thought to be powered by a BH of mass in the range 25--100 M$_{\odot}$ (\citealt{2014MNRAS.439L...1C,2015MNRAS.452...24C}).

The presence of IMBHs was also suggested in the HLXs 2XMM J011942.7+032421 (\citealt{2012MNRAS.423.1154S}; \citealt{2014ApJ...797L...7G}) and CXO J122518.6+144545 (\citealt{2010MNRAS.407..645J}; \citealt{2015MNRAS.454L..26H}), and in the ULXs NGC 1313 X-1 (e.g., \citealt{2015ApJ...811L..11P}), NGC 5408 X-1 (e.g., \citealt{2012ApJ...753..139D}), and M51 ULX-7 (\citealt{2016MNRAS.456.3840E}), among others. The HLX 2XMM J011942.7+032421, with a peak X-ray luminosity of 1.53 $\times 10^{41}$ erg s$^{-1}$, presents short-term variability and its X-ray spectrum can be fitted by multicolor disk emission with a mass $\geq$ 1900 M$_{\odot}$ (\citealt{2012MNRAS.423.1154S}). Its optical spectrum confirms the location of the HLX in the spiral arm of the galaxy NGC 470 and shows a high-ionization HeII emission line with a large velocity dispersion, which suggests the presence of a compact ($<$5 AU) highly ionized accretion disk (\citealt{2014ApJ...797L...7G}). The HLX CXO J122518.6+144545 (\citealt{2010MNRAS.407..645J}) reached a peak X-ray luminosity of 2.2 $\times 10^{41}$ erg s$^{-1}$ and its X-ray count rate varies by a factor $>60$, making it the only second outbursting HLX after HLX-1 (\citealt{2015MNRAS.454L..26H}). It also exhibits optical variability likely related to the X-ray variability. Its high X-ray luminosity makes it another strong IMBH candidate. NGC 1313 X-1 shows 3:2 ratio QPOs (\citealt{2015ApJ...811L..11P}) and has an X-ray luminosity above $10^{40}$ erg s$^{-1}$. The first X-ray spectral studies suggested that NGC 1313 X-1 hosts an IMBH of $\sim$1000 M$_{\odot}$ (\citealt{2003ApJ...585L..37M,2013ApJ...776L..36M}). A later analysis and applying the same scaling as for M82 X-1 yielded a BH mass estimate for NGC1313 X-1 of 5000 $\pm$ 1300 M$_{\odot}$ (\citealt{2015ApJ...811L..11P}). However, the recent detection of a spectral cutoff above 10 keV and of an ultra-fast (0.2c) disk wind indicative of supercritical accretion rule out the IMBH scenario for this source (\citealt{2013ApJ...778..163B}; \citealt{2015MNRAS.447.3243M}; \citealt{2016Natur.533...64P}; \citealt{2016ApJ...826L..26W}). Another controversial case was that of NGC 5408 X-1. It also shows QPOs, clear variability (on scales of minutes days, months and years), and a peak X-ray luminosity of $10^{40}$ erg s$^{-1}$ (e.g., \citealt{2004A&A...423..955S}; \citealt{2007ApJ...660..580S}, \citeyear{2009ApJ...703.1386S}; \citealt{2009MNRAS.397.1061H}; \citealt{2009ApJ...702.1679K}; \citealt{2013MNRAS.435.2665C}); however, the constancy of its X-ray spectral parameters (\citealt{2009ApJ...702.1679K}) and variability of the QPO frequency (\citealt{2012ApJ...753..139D}) hampered a robust conclusion on its nature. While \cite{2012ApJ...753..139D} proposed the presence of an IMBH of at least 800 M$_{\odot}$, \cite{2011MNRAS.411..644M,2014MNRAS.438L..51M} and \cite{2015ApJ...814...73S} argued in favor of a super-Eddington accreting BH whose winds could explain the spherically-symmetric nebula observed around NGC 5408 X-1 in the radio and optical bands (\citealt{2003RMxAC..15..197P}; \citealt{2006MNRAS.368.1527S}; \citealt{2007ApJ...666...79L}; \citealt{2012ApJ...749...17C}). The recent detection of an ultra-fast outflow in the X-ray spectrum of NGC 5408 X-1 finally confirmed the super-Eddington accreting nature of this source (\citealt{2012ApJ...749...17C}). Last but not least, the ULX-7 in the spiral galaxy M51 is suggested to be powered by an IMBH based on its hard spectrum, high rms variability, and the lower limit on the BH mass of 1.6 $\times$ 10$^{3}$ M$_{\odot}$ derived using the relationship between BH mass and high frequency break in the power spectrum (\citealt{2016MNRAS.456.3840E}). However, a pulsar nature for this ULX cannot be ruled out.

The search for IMBHs in ULXs, and specially in HLXs, is an active field in which many strong IMBH candidates are being found either by cross-correlating X-ray and optical catalogs (e.g., \citealt{2012MNRAS.423.1154S}; \citealt{2016ApJ...817...88Z}) or serendipitously (e.g., \citealt{2015ApJ...814....8K}). Detailed analysis is yet required in order to distinguish between other possible scenarios proposed to explain their X-ray luminosity, such as super-Eddington accretion. 

\vspace{-0.5 cm}

\subsection{Dwarf galaxies}
\label{dwarf}
\begin{figure*}
\includegraphics[width=\textwidth]{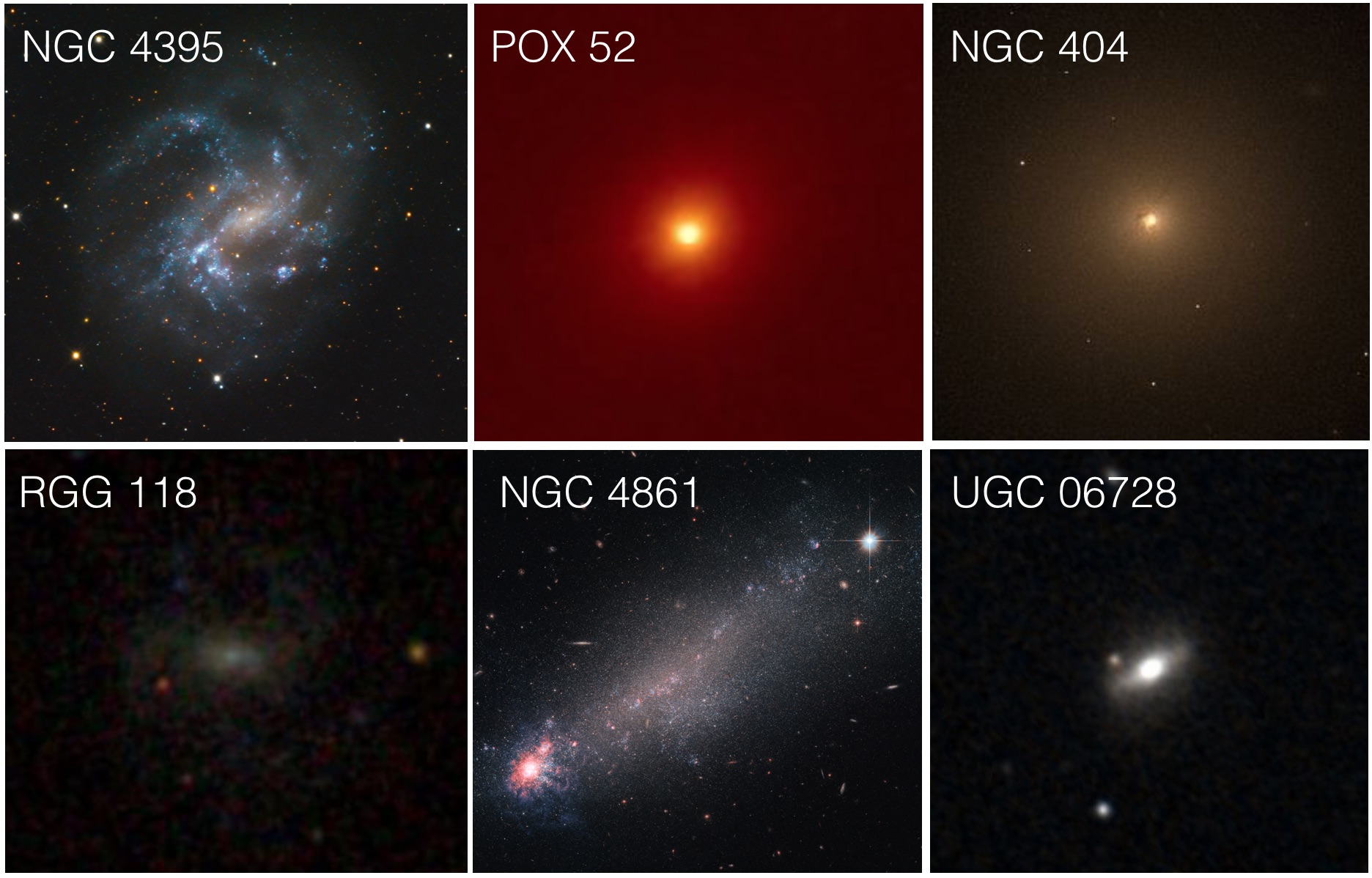}
\caption{Examples of dwarf galaxies hosting IMBHs. From left to right, top to bottom: spiral galaxy NGC 4395, image from Bob Franke, Focal Pointe Observatory; \textit{HST} image of the elliptical galaxy POX 52; \textit{HST} image of the lenticular galaxy NGC 404; SDSS image of the disk galaxy RGG 118; \textit{HST} image of the irregular galaxy NGC 4861; DSS image of the late-type galaxy UGC 06728.}
\label{mosaic}
\end{figure*}

Unlike massive galaxies, dwarf galaxies (M$_{*} \leq 3 \times 10^{9}$ M$_{\odot}$) have not significantly grown through mergers/accretion and thus resemble those galaxies formed in the early Universe. They constitute thus one of the best places where to look for seed BHs (e.g., \citealt{2010A&ARv..18..279V}; \citealt{2012NatCo...3E1304G}; \citealt{2016PASA...33...54R}). The first observational evidence for IMBHs in dwarf galaxies were identified in the late 80s in the spiral galaxy NGC\,4395 (M$_{*} \sim1.3 \times 10^{9}$ M$_{\odot}$; Fig.~\ref{mosaic}) and the elliptical galaxy Pox 52 (M$_{*} \sim1.2 \times 10^{9}$ M$_{\odot}$; Fig.~\ref{mosaic}) by the finding of high-ionization narrow emission lines and broad Balmer emission lines in their optical spectrum (\citealt{1987AJ.....93...29K}; \citealt{1989ApJ...342L..11F}). These, together with the detection of hard X-ray emission, indicate the presence of an AGN with a BH mass of $M_\mathrm{BH} \sim3 \times 10^{5}$ M$_{\odot}$ estimated from the width of the broad emission lines under the assumption that the gas is virialized (\citealt{2003ApJ...588L..13F}; \citealt{2004ApJ...607...90B}; \citealt{2005AJ....129.2108M}; \citealt{2005ApJ...632..799P}; \citealt{2005MNRAS.356..524V}; \citealt{2008ApJ...686..892T}). NGC\,4395 has, in addition, a compact radio jet (\citealt{2006ApJ...646L..95W}) and is one of the few dwarf galaxies for which a dynamical BH mass measurement has been possible ($M_\mathrm{BH} = 4^{+8}_{-3} \times 10^{5}$ M$_{\odot}$; \citealt{2015ApJ...809..101D}). An upper limit on the BH mass of $M_\mathrm{BH} = 1.5 \times 10^{5}$ M$_{\odot}$ was also estimated from dynamical modeling for the dwarf S0 galaxy NGC\,404 (\citealt{2010ApJ...714..713S}; \citealt{2017ApJ...836..237N}) and is in agreement with that derived using the fundamental plane of BH accretion (\citealt{2012ApJ...753..103N}; but see \citealt{2014ApJ...791....2P}). Further evidence for the presence of an accreting BH in NGC\,404 comes from its optical classification as a LINER\footnote{Low-ionization nuclear emission line region (LINERs) are associated with low-luminosity AGN (e.g., \citealt{2005A&A...435..521N}; \citealt{2014ApJ...787...62M})}, and from the finding of a hard X-ray core (\citealt{2011ApJ...737...77B}), UV variability (\citealt{2005ApJ...625..699M}), unresolved and variable hot dust emission (\citealt{2010ApJ...714..713S}), and mid-infrared (MIR) AGN-like emission lines (\citealt{2004A&A...414..825S}). 

The detection of hard X-ray emission spatially coincident with core radio emission is, in the absence of dynamical mass measurements, a very strong tracer of BH accretion. This yielded the discovery of the first AGN in a blue compact dwarf galaxy (Henize 2-10; \citealt{2011Natur.470...66R}). Although Henize 2-10 is a starburst galaxy whose optical emission is dominated by star formation, it hosts a low-mass BH ($M_\mathrm{BH} \sim10^{6}$ M$_{\odot}$) at its center as revealed by \textit{Chandra} X-ray observations and VLA and VLBI radio observations (\citeauthor{2011Natur.470...66R} 2011, \citeyear{2016ApJ...830L..35R}; \citealt{2012ApJ...750L..24R}). A low-mass BH was also found based on spatially coincident \textit{Chandra} X-ray emission and VLA radio emission in Mrk 709, a low-metallicity blue compact dwarf formed by a pair of interacting galaxies (\citealt{2014ApJ...787L..30R}). Mrk 59, the compact core of the blue compact dwarf galaxy NGC 4861, has been also found to host an IMBH of $\sim5 \times 10^{4}$ M$_{\odot}$ based on high-resolution X-ray and radio observations with \textit{Chandra} and the European VLBI Network, respectively, and the detection of high-excitation emission lines (e.g., HeII) typically associated with gas photoionized by AGN (\citealt{2011AN....332..379M}; Yang et al. in preparation)

\begin{figure*}[b]
\includegraphics[width=\textwidth]{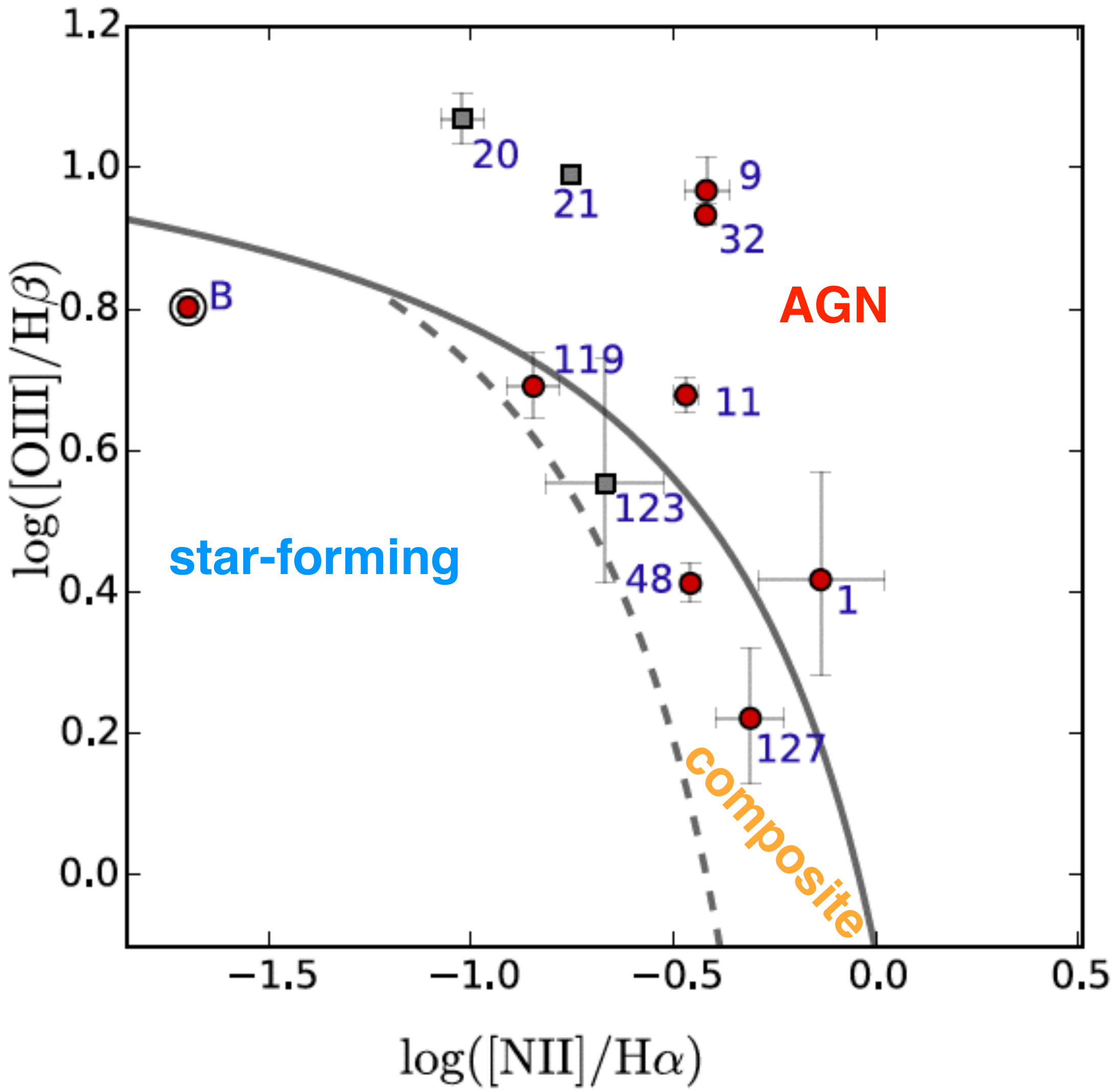}
\caption{Location of the 10 dwarf galaxies from \cite{2013ApJ...775..116R} detected in X-rays on the narrow-line emission diagnostic diagram. Red circles represent sources with \textit{Chandra} observations from \cite{2017ApJ...836...20B}, gray squares represent objects with archival \textit{Chandra} observations. Those sources that qualify as broad-line AGN or composite have X-ray luminosities significantly higher than would be expected from star formation. The only exception is source B, which qualifies as star-forming according to the diagram and whose X-ray luminosity is consistent with that from XRBs. Figure and caption adapted from \cite{2017ApJ...836...20B}. \copyright AAS. Reproduced with permission.}
\label{BPT}
\end{figure*}

The first searches for low-mass BHs had already begun with the arrival of the Sloan Digital Sky Survey (SDSS), which provided optical spectra for more than 100,000 galaxies. This allowed a systematic search for low-mass BHs through the kinematics and the ionization properties of the excited gas: the broad-line widths of the gas provide an estimate of the BH mass under the assumption that the gas is virialized, while narrow emission line diagnostics (e.g.,  [OIII]/H$\beta$ versus [NII]/H$\alpha$; \citealt{2003MNRAS.346.1055K}; \citealt{2006MNRAS.372..961K}) are used to distinguish between AGN and starburst emission (\citealt{2004ApJ...610..722G,2005ApJ...630..122G,2007ApJ...670...92G}; \citealt{2007ApJ...657..700D}; \citealt{2008AJ....136.1179B}; \citealt{2012ApJ...755..167D}). This yielded the identification of 229 (\citealt{2004ApJ...610..722G,2007ApJ...670...92G}) and 309 (\citealt{2012ApJ...755..167D}) low-mass AGN with BH masses $< 2 \times 10^{6}$ M$_{\odot}$. Most of the host galaxies of these low-mass BHs are of late-type and more massive than typical dwarf galaxies. It was the discovery of Henize 2-10 that invigorated the quest for low-mass AGN in dwarf galaxies: \cite{2013ApJ...775..116R} found 136 optically selected dwarf galaxies at $z < 0.055$ that qualified as either AGN or composite objects in the narrow-line emission diagnostic diagram (see Fig.~\ref{BPT}) and provided an AGN fraction (not corrected for incompleteness) of 0.5\%. Ten of these sources present broad optical emission lines, from which a range of BH masses of $\sim7 \times 10^{4}-1 \times 10^{6}~M_{\odot}$ was estimated, and have X-ray luminosities significantly higher than what would be expected from star formation, thus confirming the presence of accreting BHs in these galaxies (\citealt{2017ApJ...836...20B}). Follow-up studies of the source RGG 118 in \cite{2013ApJ...775..116R}, a dwarf galaxy classified as a composite object and for which there was a hint of broad emission in the SDSS spectrum, revealed the presence of hard X-ray emission and broad H$\alpha$ line emission, from which a BH mass of $\sim5 \times 10^{4}$ M$_{\odot}$ was estimated (\citealt{2015ApJ...809L..14B}). This makes RGG 118 the lightest nuclear BH known.
Using also optical emission line diagnostics and SDSS data, \cite{2014AJ....148..136M} identified 18 additional IMBH candidates with a minimum BH mass in the range $10^{3}-10^{4}$ M$_{\odot}$. Most of their host galaxies have stellar masses above the typical threshold of $3 \times 10^{9}$ M$_{\odot}$. \cite{2015MNRAS.454.3722S} identified 3 additional AGN candidates using a combined criterion that includes MIR color cuts in addition to the classical narrow-line diagnostic diagram. MIR color searches rely on the different colors of dust when heated by AGN or by stars or non-active galaxies and have become a very common tool for identifying AGN, specially since the arrival of the \textit{Wide-Field Infrared Survey Explorer} (\textit{WISE}; e.g., for the \textit{WISE} bands $W1$ and $W2$ at 3.4$\mu$m and 4.6$\mu$m, respectively, AGN can be identified as having $W1-W2 \geq 0.8$, \citealt{2012ApJ...753...30S}). Although several studies have made use of MIR colors cuts for selecting AGN in low-mass galaxies (e.g., \citealt{2014ApJ...784..113S}; \citealt{2014arXiv1411.3844M}), caution should be taken when using this selection technique as star-forming dwarf galaxies can show similar MIR colors to those of luminous AGN (\citealt{2016ApJ...832..119H}). Other MIR searches are based on the detection of the high-ionization emission line [NeV] 14 $\mu$m or the 24 $\mu$m line using \textit{Spitzer} spectral observations. This yielded the detection of 9 AGN in bulgeless or late-type (with a minimal bulge) galaxies, for which a lower limit on the BH mass ranging from $\sim3 \times 10^{3}-1.5 \times 10^{5}~M_{\odot}$ was estimated assuming sub-Eddington accretion (\citeauthor{2007ApJ...663L...9S} 2007, \citeyear{2008ApJ...677..926S}, \citeyear{2009ApJ...704..439S}). Follow-up observations with the \textit{XMM-Newton} X-ray satellite revealed the presence of hard, unresolved X-ray emission in one of these bulgeless dwarf galaxies (J1329+3234), with an X-ray luminosity ($L_\mathrm{X} = 2.4 \times 10^{40}$ erg s$^{-1}$) two orders of magnitude larger than that expected from star formation and consistent with an accreting BH (\citealt{2015ApJ...798...38S}). NGC 4178 is another late-type bulgeless disk galaxy that has a prominent [NeV] emission line suggesting the presence of an AGN (\citealt{2009ApJ...704..439S}). \textit{Chandra} observations of NGC 4178 reveal the presence of unresolved nuclear X-ray emission spatially coincident with the dynamical center of the galaxy and suggest that the AGN is heavily absorbed and accreting at high rates (\citealt{2012ApJ...753...38S, 2013ApJ...777..139S}). Using the fundamental plane of accreting BHs, the correlation between nuclear stellar cluster mass and BH mass, and the bolometric luminosity, the authors estimate a range of BH masses for this source of $\sim10^{4}-10^{5}$ M$_{\odot}$. Using reverberation mapping, \cite{2016ApJ...831....2B} found another low-mass BH in a nearby late-type galaxy, UGC 06728, which hosts a low-luminosity Seyfert 1 AGN with a BH mass of $M_\mathrm{BH} = (7.1 \pm 4.0) \times 10^{5}$ M$_{\odot}$

\begin{figure*}
\includegraphics[width=\textwidth]{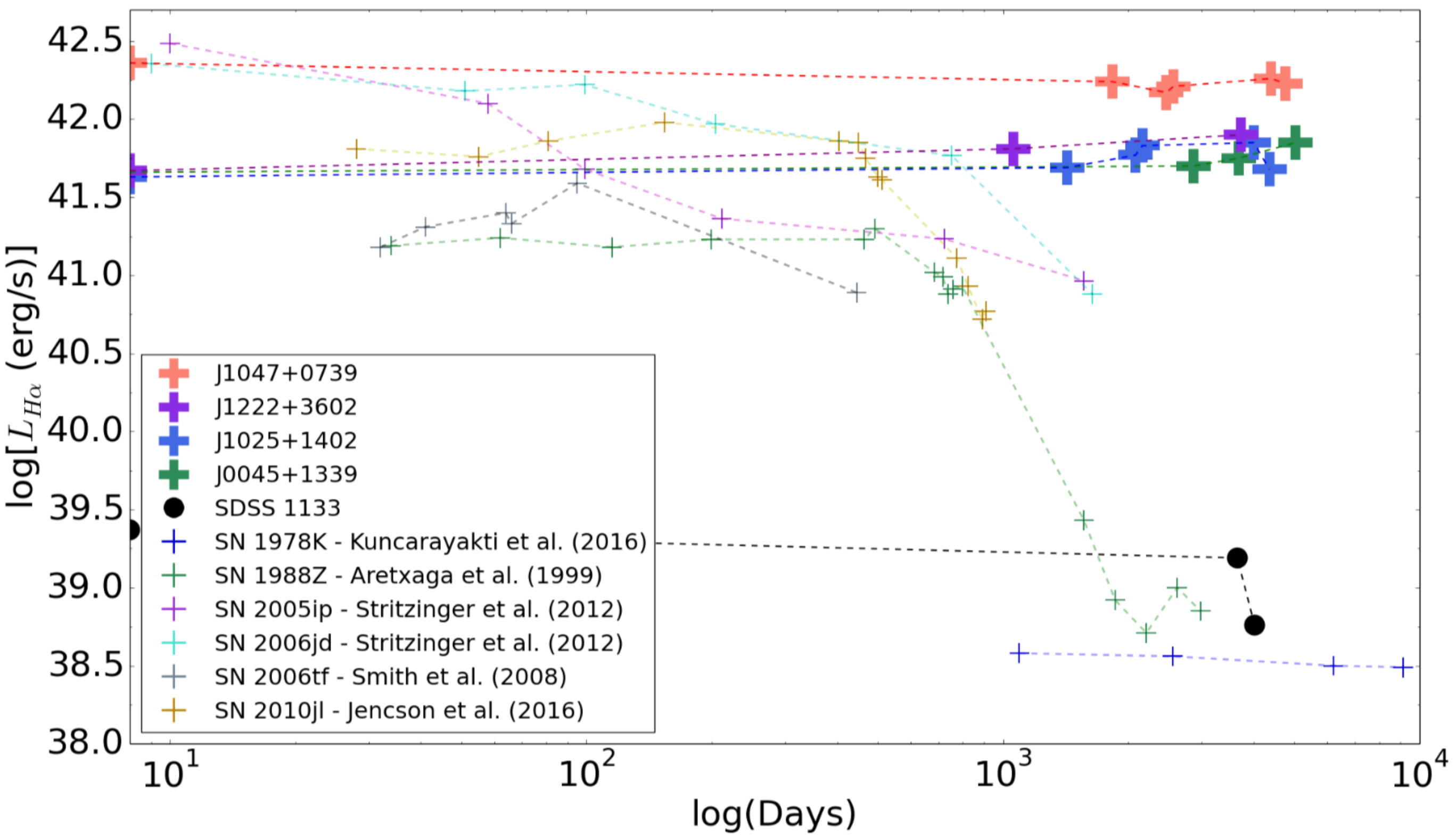}
\caption{Time evolution of the broad H$\alpha$ luminosity for the low-metallicity dwarf galaxies of \cite{2016A&A...596A..64S} as well as several luminous type IIn supernovae and the transient event SDSS 1133 (\citealt{2014MNRAS.445..515K}). The low-metallicity dwarf galaxies are roughly constant over periods of 10-13 yr. Error bars are smaller than the symbols. Figure and caption from \cite{2016A&A...596A..64S}, A\&A reproduced with permission. \copyright ESO.}
\label{Simmonds}
\end{figure*}

The detection of broad H$\alpha$ and H$\beta$ emission lines so commonly used to estimate BH masses should not be used as the only tool for identifying AGN, as the emission of broad optical lines might come from transient stellar processes rather than the AGN ionized gas. Evidence for this is provided by the finding that the broad H$\alpha$ emission of those objects from \cite{2013ApJ...775..116R} classified as star-forming in the narrow-line diagnostic diagram has faded over a time range of 5-14 years, while those falling in the AGN region of the diagram present persistent broad H$\alpha$ emission (\citealt{2016ApJ...829...57B}). An intriguing case is that found by \cite{2016A&A...596A..64S}, who studied a sample of low-metallicity dwarf galaxies with broad H$\alpha$ emission: although their sources present long-lived ($> 10$ yr) broad H$\alpha$ emission lines incompatible with a supernova origin (see Fig.~\ref{Simmonds}), they lack the strong X-ray emission and non-thermal hard UV emission characteristic of AGN. This implies that these sources are either a particular case of AGN with very weak X-ray and UV emission, with fully obscured accretion disks, or that they are not AGN and the persistent broad emission lines are produced by very extreme stellar processes. 

\begin{figure*}
\includegraphics[width=\textwidth]{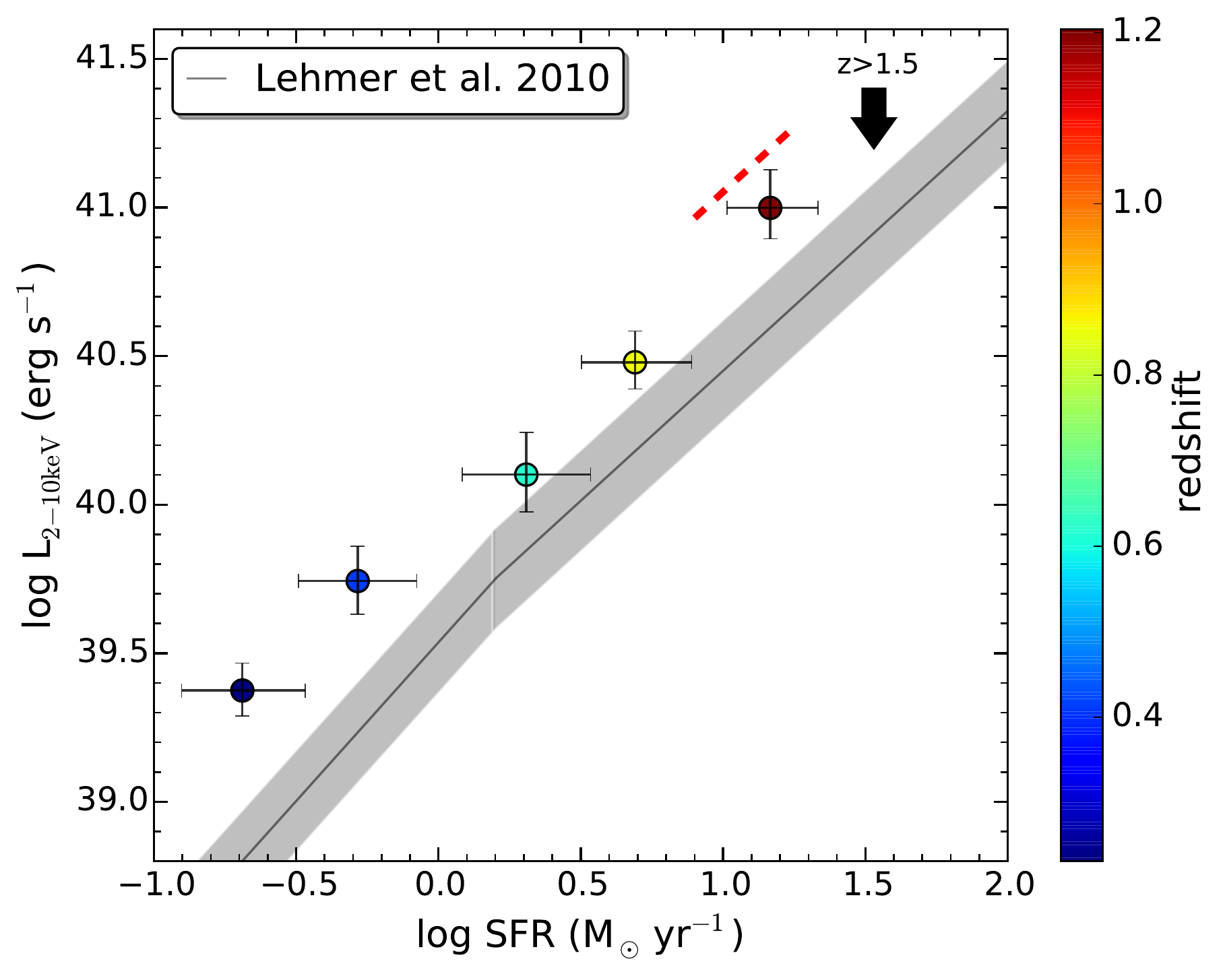}
\caption{2-10 keV X-ray luminosity vs. average star-formation rate (SFR) of each stacked redshift bin for the sample of $\sim$50,000 starburst and late-type dwarf galaxies of \cite{2016ApJ...817...20M}. The gray line shows the correlation between SFR, stellar mass and XRB luminosity from \cite{2010ApJ...724..559L} with a scatter of 0.34 dex. The 1$\sigma$ error bars account for the stacking uncertainties and the statistical errors on the SFRs and stellar masses. The red dashed line indicates where the expected contribution from XRBs would lie if the metallicity were a factor three lower than the solar one at z $>$ 1 as predicted by \cite{2013ApJ...776L..31F}. The stacked X-ray emission is more than 3$\sigma$ higher than the one expected from XRBs for the five complete redshift bins. Figure and caption from \cite{2016ApJ...817...20M}. \copyright AAS. Reproduced with permission.}
\label{stacking}
\end{figure*}

Unlike optical studies, which are often skewed toward high Eddington ratios, X-ray searches offer the advantage of probing low rates of accretion and detecting AGN so faint as XRBs ($L_\mathrm{X}\sim10^{38}$ erg s$^{-1}$; e.g., \citealt{2010ApJ...714...25G}). X-ray surveys also cover larger volumes, and make possible the detection of IMBHs in dwarf galaxies at intermediate redshifts (Mezcua et al. in preparation), at an epoch when cosmic star formation and AGN activity reached their peak (\citealt{1999MNRAS.310L...5F}; \citealt{2005ApJ...624..630S}). One of the first searches for accreting BHs in low-mass galaxies that made use of deep \textit{Chandra} and \textit{XMM-Newton} X-ray surveys is that of \cite{2008ApJ...688..794S}, who found 32 objects out to $z \sim1$. The authors identified low-mass galaxies as having M$_{*} < 2 \times 10^{10}$ M$_{\odot}$ and considered only sources with $L_\mathrm{X} > 10^{42}$ erg s$^{-1}$, hence their sample is formed by SMBHs with an average BH mass of $3 \times 10^{6}$ M$_{\odot}$. Using the AGN Multiwavelength Survey of Early-Type Galaxies (AMUSE; in the Virgo cluster, 454 ks, \citealt{2008ApJ...680..154G}; in the field within 30 Mpc, 479 ks, \citealt{2012ApJ...747...57M}), \cite{2015ApJ...799...98M} found nuclear X-ray emission possibly coming from accreting BHs in 7 early-type galaxies with $M_{*} < 10^{10}$ M$_{\odot}$ and obtain a lower limit on the BH occupation fraction of $> 20\%$. Using a stacking analysis of early-type galaxies in the \textit{Chandra} COSMOS survey (0.9 deg$^{2}$, 1.8 Ms; \citealt{2009ApJS..184..158E}; \citealt{2012ApJS..201...30C}) and after removing the contribution from XRBs and hot gas to the X-ray emission, \cite{2016ApJ...823..112P} also found that highly absorbed AGN are present in low-mass early-type galaxies, with BH masses ranging from $10^{6} - 10^{8}$ M$_{\odot}$. \cite{2016ApJ...817...20M} applied the same stacking technique to a sample of $\sim$50,000 starburst and late-type dwarf galaxies up to redshift $\sim$1.5 in the \textit{Chandra} COSMOS Legacy survey (2.2 deg$^{2}$, 4.6 Ms; \citealt{2016ApJ...819...62C}), finding also an X-ray excess that can be explained by accreting BHs with $M_\mathrm{BH} \sim10^{5}$ M$_{\odot}$ and X-ray luminosities as low as 10$^{39}$ erg s$^{-1}$ (see Fig.~\ref{stacking}). The authors concluded that a population of IMBHs exists in dwarf starburst galaxies but that their detection beyond the local Universe is challenging due to their low luminosity and mild obscuration. Yet, the use of wide-area X-ray surveys such as COSMOS Legacy provide one of the best tools for detecting IMBHs at intermediate redshifts: Mezcua et al. (in preparation) find 47 dwarf galaxies in the COSMOS Legacy with AGN X-ray luminosities $L_\mathrm{0.5-10 keV}$ ranging $\sim4 \times 10^{39}$ erg s$^{-1}$ to $10^{44}$ erg s$^{-1}$ and redshifts as high as $z \sim$2.4. The BH masses range from $\sim10^{4}$ M$_{\odot}$ to $\sim8 \times 10^{5}$ M$_{\odot}$, indicating that all the sources likely host an IMBH. This constitutes the largest sample of IMBHs beyond the local Universe so far discovered. 

Making use of the 4 Ms \textit{Chandra} Deep Field South (CDF-S; \citealt{2001ApJ...551..624G,2002ApJS..139..369G}) and also applying the stacking technique, \cite{2012ApJ...758..129X} found that the unresolved 6-8 keV cosmic X-ray background is mostly produced by low-mass galaxies with obscured AGN at $z \sim1-3$. The discovery of three individual IMBHs with $M_\mathrm{BH} \sim2 \times 10^{5}$ M$_{\odot}$ in dwarf galaxies at $z < 0.3$ in the Extended CDF-S (0.3 deg$^{2}$) was reported by \cite{2013ApJ...773..150S}. \cite{2016ApJ...831..203P} found ten more individual detections up to $z < 0.6$ in the area of the All-Wavelength Extended Groth Strip International Survey (AEGIS) field ($\sim$200--800 ks; \citealt{2007ApJ...660L...1D}) covered by \textit{Chandra} (0.1 deg$^{2}$), and derived an AGN fraction for dwarf galaxies with $10^{9} < M_{*} < 3 \times 10^{9}$ M$_{\odot}$ at 0.1 $< z <$ 0.6 of $\sim3 \%$. \cite{2015ApJ...805...12L} found that 19 out of $\sim$44,000 dwarf galaxies in the NASA-Sloan Atlas\footnote{\url{http://www.nsatlas.org}} have \textit{Chandra} hard X-ray detections. Eight of these sources present enhanced X-ray emission with respect to that of star formation and are potential IMBH candidates with $L_\mathrm{X} \sim10^{38}-10^{40}$ erg s$^{-1}$, while the X-ray luminosity of the rest of the sources can be explained by XRBs. Using also the NASA-Sloan Atlas, \cite{2017ApJ...837...66N} identified 19 low-mass galaxies (with stellar masses up to 10$^{10}$ M$_{\odot}$) with \textit{XMM-Newton} X-ray emission and radio emission spatially coincident with the galaxy center. Using the fundamental plane of BH accretion, the authors derive a range of BH masses of 10$^{4}-2 \times 10^{8}$ M$_{\odot}$. A few more tens of IMBH candidates have been found using pointed observations. \textit{Chandra} observations of 66 of the 229 low-mass AGN identified by \cite{2004ApJ...610..722G,2007ApJ...670...92G} revealed the detection of X-ray emission in 52 sources and confirmed their AGN nature (\citealt{2009ApJ...698.1515D}; \citealt{2012ApJ...761...73D}; \citealt{2014ApJ...788L..22G}; \citealt{2014ApJ...782...55Y}; \citealt{2016ApJ...825..139P}). \cite{2011ApJ...731...55J} carried out \textit{XMM-Newton} observations of six Lyman Break Analogs, which are the local analogs to the high-redshift star-forming Lyman-Break galaxies\footnote{Lyman-Break galaxies are high redshift ($z > 6$) massive galaxies that are expected to host a BH by that time (e.g., \citealt{2010A&ARv..18..279V}).}. The intermediate starburst-type 2 AGN classification of their optical emission line spectra, the detection of hard X-ray emission with $L_\mathrm{X} \sim10^{42}$ erg s$^{-1}$, and MIR to [OIII] luminosity ratios higher than those of type 2 AGN indicate the possible presence of low-mass AGN in these targets, with a BH mass ranging between $10^{5} - 10^{6}$ M$_{\odot}$ assuming the sources radiate at the Eddington limit. Based on rapid X-ray variability, \cite{2012ApJ...751...39K} presented an additional sample of 15 low-mass AGN candidates, 7 of which have $M_\mathrm{BH} < 2 \times10^{6}$ M$_{\odot}$ and are thus candidates to IMBHs. Optical spectroscopy of 12 of these low-mass BHs revealed the presence of broad H$\alpha$ emission lines, from which BH masses in the range $10^{5} - 10^{6}$ M$_{\odot}$ were estimated (\citealt{2016ApJ...821...48H}). 


\subsubsection*{\textbf{X-ray weak IMBHs: hiding behind the dust?}} 
\begin{figure*}
\includegraphics[width=\textwidth]{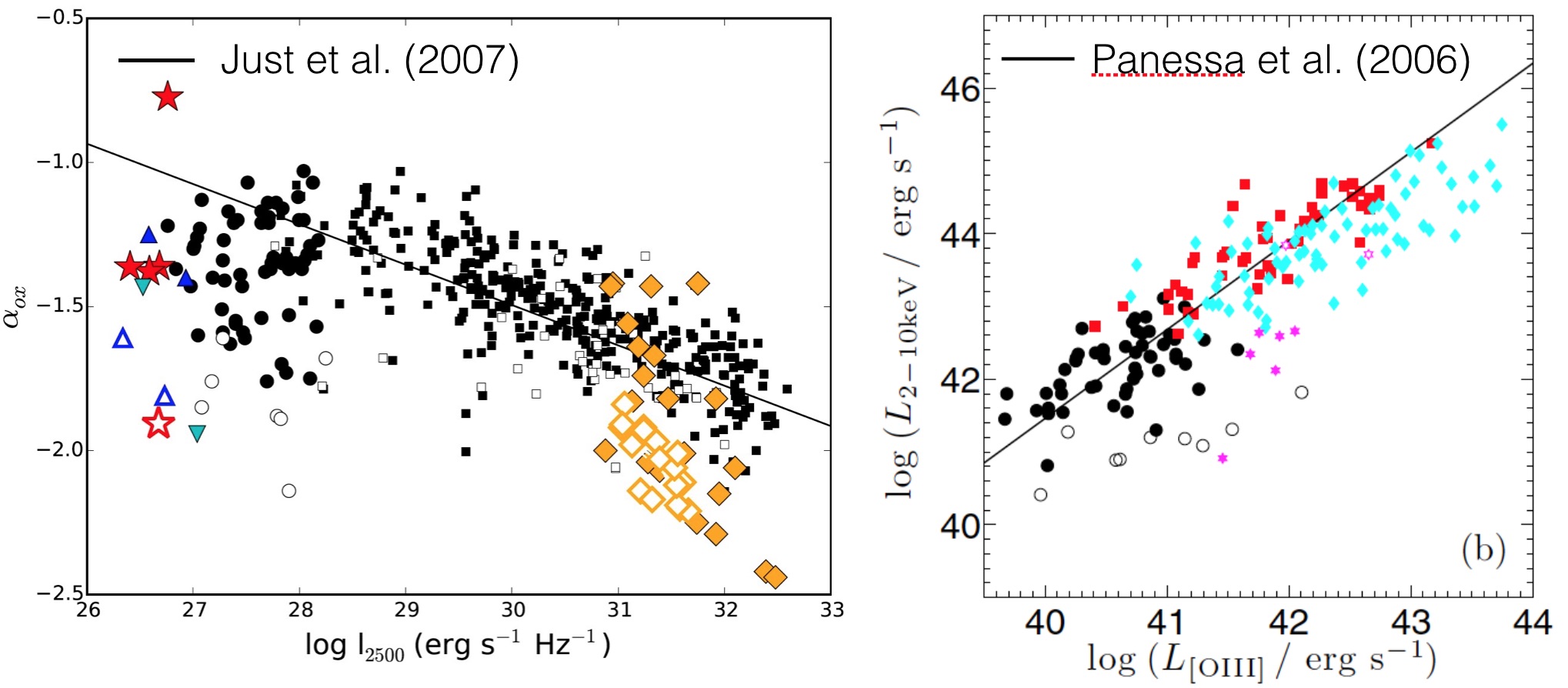}
\caption{\textbf{Left:} $\alpha_\mathrm{OX}$ vs. $l_\mathrm{2500}$ for the low-mass AGN from the \cite{2004ApJ...610..722G,2007ApJ...670...92G} sample observed by \textit{Chandra} by \citeauthor{2016ApJ...825..139P} (2016; red stars, sources with low Eddington ratios), \citeauthor{2014ApJ...782...55Y} (2014; blue triangles, sources with low Eddington ratios), \citeauthor{2014ApJ...788L..22G} (2014; cyan upside down triangles, sources with low Eddington ratios), and \citeauthor{2012ApJ...761...73D} (2012; circles, sources with high Eddington ratios). SMBHs are plotted for comparison, including weak emission line quasars from \cite{2012ApJ...747...10W} and \citeauthor{2015ApJ...805..122L} (2015; orange diamonds) and 'normal' type 1 quasars from \citeauthor{2007ApJ...665.1004J} (2007; squares). \textbf{Right:} L$_\mathrm{2-10 keV}$ vs. L$_\mathrm{[OIII]}$ for the low-mass AGN from \citeauthor{2012ApJ...761...73D} (2012; circles). Optical quasars from \citeauthor{2012MNRAS.425..907J} (2012; red squares) and quasars from \citeauthor{2000ApJ...528..637B} (2000; light blue diamonds; X-ray weak objects shown in magenta hexagrams) are plotted for comparison. All open symbols denote no detections. Figures and caption adapted from \cite{2012ApJ...761...73D} and \cite{2016ApJ...825..139P}. \copyright AAS. Reproduced with permission.}
\label{Xraytooptical}
\end{figure*}

Both for the \cite{2004ApJ...610..722G,2007ApJ...670...92G} sample and the \cite{2013ApJ...775..116R} sample studied in X-rays by \cite{2017ApJ...836...20B}, the X-ray observations show that most of the low-mass AGN have X-ray to UV ratios ($\alpha_\mathrm{OX}$) below the correlation between $\alpha_\mathrm{OX}$ and luminosity density at 2500 $\AA$ ($l_\mathrm{2500}$) defined for more luminous sources hosting SMBHs (\citealt{2007ApJ...665.1004J}; see Fig.~\ref{Xraytooptical}, left). This behavior was also found for broad absorption-line quasars whose X-ray weakness is caused by absorption (e.g., \citealt{2000ApJ...528..637B}; \citealt{2001ApJ...546..795G}). The puzzling X-ray weak tail of low-mass AGN does not seem to be caused by differences between low- and high-Eddington rates nor slim disk accretion (\citealt{2014ApJ...782...55Y}; \citealt{2016ApJ...825..139P}), while variability would scatter the sources around the correlation instead of them lying systematically below it. Are then these low-mass AGN intrinsically X-ray weak? Or are their accretion disks fully obscured along our line-of-sight? The latter scenario is supported by the [OIII] deficit of some of the low-mass AGN compared to their X-ray luminosity (i.e., they lie below the relation between 2-10 keV X-ray luminosity and [OIII]$\lambda$5007 optical line luminosity defined by higher luminosity unobscured AGNs; see Fig.~\ref{Xraytooptical}, right), a behavior typical of heavily obscured AGN (e.g., \citealt{2006A&A...455..173P}). Similar results were found by Simmonds et al. (2016), whose low-mass AGN lie at least $\sim$1-2 dex below the relation between [OIII]$\lambda$5007 luminosity and 2-10 keV luminosity. Hard X-ray ($\geq$10 keV) observations with \textit{NuSTAR} could test whether these weak low-mass AGN are indeed obscured. \cite{2017ApJ...837...48C} searched for low-mass AGN using the 40-month \textit{NuSTAR} serendipitous survey, finding 10 sources with median stellar mass $<M_{*}> = 4.6 \times 10^{9}$ M$_{\odot}$, X-ray luminosity $<L_\mathrm{2-10 keV}> = 3.1 \times 10^{42}$ erg s$^{-1}$ and BH masses in the range (1.1 - 10.4) $\times 10^{6}$ M$_{\odot}$. Although 30\% of their sources do not show AGN-like optical narrow emission lines, only one source is found to be heavily obscured (with a column density $N_\mathrm{H} > 10^{23}$ cm$^{-2}$). Similar results were found by \cite{2015MNRAS.447.2112L}, who studied \textit{XMM-Newton} data of 14 low-mass AGN drawn from the \cite{2007ApJ...670...92G} sample and found that only two of them show evidence for significant absorption. 

\subsubsection*{\textbf{Do IMBHs follow the trends?}}
\begin{figure*}[h!]
\includegraphics[width=\textwidth]{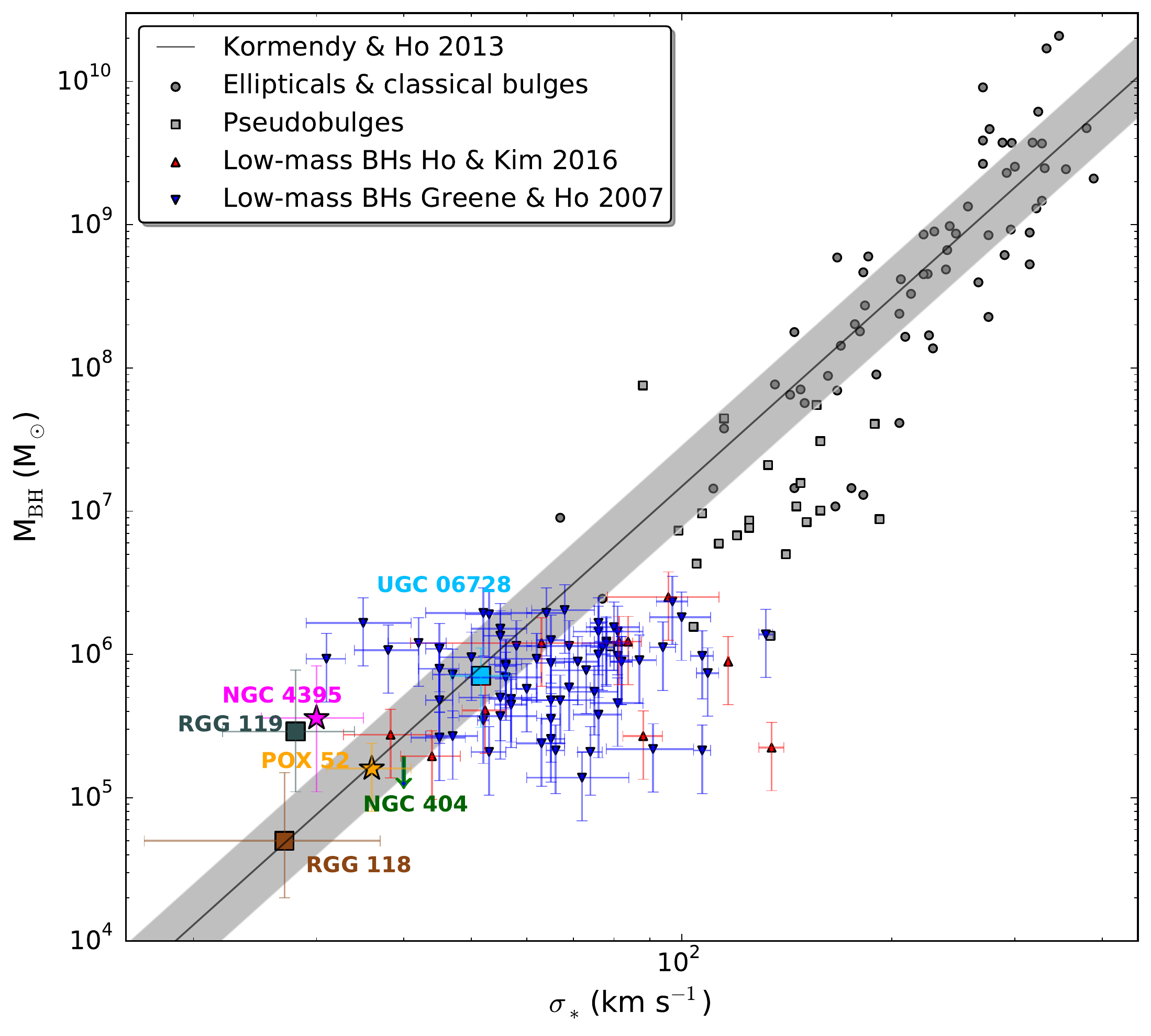}
\caption{Relation between BH mass and stellar velocity dispersion for inactive BH with dynamical mass measurements compiled by \citeauthor{2013ARA&A..51..511K} (2013; gray circles show ellipticals and classical bulges; grey squares show pseudobulges), and for low-mass AGN with virial BH masses measured using the reverberation mapping technique (UGC 06728, \citealt{2016ApJ...831....2B}; red triangles show the sample of \citealt{2016ApJ...821...48H}; inverted blue triangles the sample of \citealt{2007ApJ...670...92G}). The IMBHs NGC 4395 (\citealt{2003ApJ...588L..13F}; \citealt{2005ApJ...632..799P}), POX 52 (\citealt{2004ApJ...607...90B}), NGC 404 (\citealt{2002AJ....124.2607B}; \citealt{2017ApJ...836..237N}), RGG 118 (\citealt{2015ApJ...809L..14B}), and RGG 119 (\citealt{2017ApJ...836...20B}) are also shown. The solid line shows the $M_\mathrm{BH}-\sigma_{*}$ relation for ellipticals and classical bulges from \cite{2013ARA&A..51..511K} with a 1$\sigma$ scatter (shaded area).}
\label{Msigma}
\end{figure*}

Whether or not IMBHs follow the BH-host galaxy correlations found for spheroidal galaxies hosting SMBHs is directly connected to the formation of seed BHs: the low-mass end of the $M_\mathrm{BH}-\sigma_{*}$ correlation (or also of the $M_\mathrm{BH}-M_\mathrm{*}$ relation; \citealt{2015ApJ...813...82R}) is expected to flatten towards what looks like a plume of ungrown BHs if seed BHs are massive (i.e. formed from direct collapse), while this asymptotic flattening or plume would lie at BH masses that cannot be currently measured observationally (e.g., \citealt{2010MNRAS.408.1139V}; \citealt{2010A&ARv..18..279V}; see Fig.~\ref{occupationfraction}) if seed BHs are light (Pop III seeds). 
In Table~\ref{sigmas} I compile all the IMBHs and low-mass AGN with $M_\mathrm{BH} \lesssim 10^{6}$ M$_{\odot}$ for which a stellar velocity dispersion is available, and plot them on the $M_\mathrm{BH}$ versus $\sigma_{*}$ diagram in Fig.~\ref{Msigma}: a tentative plume similar to that expected from massive BH seeds (Fig.~\ref{occupationfraction}, left) seems to be observed at BH masses $\sim10^{5}-10^{6}$ M$_{\odot}$, which suggests that the direct collapse scenario is the main formation mechanism of seed BHs. However, this could just be a bias caused by the easier detection of massive (i.e. of $\sim10^{5}$ M$_{\odot}$) than light (i.e. of $\sim10^{3}$ M$_{\odot}$) BHs. 

The virial BH masses of the 10 low-mass AGN from \cite{2016ApJ...821...48H} are found to follow the $M_\mathrm{BH}-\sigma_{*}$ relation (if the considerably large scatter in the pseudobulge regime is considered, e.g., \citealt{2016ApJ...821...48H}; grey squares in Fig.~\ref{Msigma}), as found for other samples of low-mass AGN (also included in Table~\ref{sigmas} and Fig.~\ref{Msigma}; e.g., \citealt{2005ApJ...619L.151B}; \citealt{2006ApJ...641L..21G}; \citealt{2011ApJ...739...28X}; \citealt{2016ApJ...831....2B}). IMBHs tend also to sit on the extrapolation of the $M_\mathrm{BH}-\sigma_{*}$ relation for early-type galaxies or AGN (\citealt{2015ApJ...809L..14B,2017ApJ...836...20B}), indicating that the $M_\mathrm{BH}-\sigma_{*}$ relation extends over five orders of magnitude in BH mass (see Fig.~\ref{Msigma}). However, this is not the case for the $M_\mathrm{BH}-M_\mathrm{bulge}$ relation: the BH mass of IMBHs/low-mass AGN tends to be lower at a given bulge mass than expected from an extrapolation of the relation found for classical bulges (e.g., \citealt{2008ApJ...688..159G}; \citealt{2011ApJ...742...68J};  \citealt{2013ApJ...764..151G,2015ApJ...798...54G}; \citealt{2015ApJ...809L..14B,2017ApJ...836...20B}; \citealt{2017ApJ...836..237N}). SMBHs in inactive galaxies with pseudobulges (\citealt{2008MNRAS.386.2242H}; \citealt{2013ARA&A..51..511K}) and spiral galaxies with megamaser BH mass measurements (\citealt{2010ApJ...721...26G}; \citealt{2016ApJ...825....3L}) are also found to fall below the $M_\mathrm{BH}-M_\mathrm{bulge}$ relation of early-type galaxies (\citealt{2013ARA&A..51..511K}), which suggests that most IMBHs/low-mass AGN are hosted by late-type galaxies (\citealt{2007ApJ...670...92G}; \citealt{2016ApJ...821...48H}). Finally, low-mass late-type galaxies tend to lie as well below the $M_\mathrm{BH}-M_\mathrm{*}$ relation compared to bulge-dominated and elliptical (early-type) galaxies (\citealt{2015ApJ...813...82R}), which could be explained by IMBH quenching of the star formation during an early gas-rich phase in the life of the dwarf galaxy (\citealt{2017arXiv170308553S}). The lower normalization found by \cite{2015ApJ...813...82R}, if it holds at high redshift, could explain the dearth of BH detections at $z > 6$ (e.g., \citealt{2011MNRAS.417.2085V}; \citealt{2015ApJ...813...82R}; see next section).



\begin{table}[t!]
\tbl{IMBHs and low-mass AGN with reported stellar velocity dispersions}
{\begin{tabular}{@{}lccc@{}} \toprule
Name		 			&  $M_\mathrm{BH}$ $^{a}$		&	$\sigma$								&	References												\\
						&	[M$_{\odot}$]  					&        [km s$^{-1}$]  						&  															\\ \colrule
NGC 404					&	$< 1.5 \times 10^{5}$			&	40 $\pm$ 3						& 		(1, 2) 			\\
NGC 4395				&	(3.6 $\pm$ 1.1) $\times 10^{5}$		&	30 $\pm$ 5						& 		(3, 4)				\\
POX 52					&	$1.6 \times 10^{5}$				&	36 $\pm$ 5						&		(5) 				\\
RGG118					&	$5 \times 10^{4}$				&	27$^{+12}_{-10}$ $^{b}$				&	 	(6) 				\\
RGG119					&	(2.9 $\pm$ 0.6) $\times 10^{4}$		&	28 $\pm$ 6						& 		(7)				\\ 
UGC 06728				&	(7.1 $\pm$ 4.0) $\times 10^{5}$		&	51.6 $\pm$ 4.9						&		(8)				\\ 
SDSS J000111.15-100155.5	&	$1.0 \times 10^{6}$				& 	76 $\pm$ 3 						&             (9, 10)			\\  
SDSS J002228.36-005830.6	&	$0.5 \times 10^{6}$				& 	57 $\pm$ 4 						&             (9, 10)			\\  
SDSS J004042.10-110957.6	&	$1.3 \times 10^{6}$				& 	55 $\pm$ 3 						&             (9, 10)			\\  
SDSS J010712.03+140844.9	&	$1.1 \times 10^{6}$				& 	38 $\pm$ 4 						&             (9, 10)			\\  
SDSS J012055.92-084945.4	&	$1.9 \times 10^{6}$				& 	53 $\pm$ 4 						&             (9, 10)			\\  
SDSS J015804.75-005221.9	&	$0.8 \times 10^{6}$				& 	45 $\pm$ 3 						&             (9, 10)			\\  
SDSS J022849.51-090153.7	&	$0.2 \times 10^{6}$				& 	63 $\pm$ 5 						&             (9, 10)			\\  
SDSS J023310.79-074813.3	&	$1.0 \times 10^{6}$				& 	107 $\pm$ 4 						&             (9, 10)			\\  
SDSS J024402.24-091540.9	&	$1.5 \times 10^{6}$				& 	76 $\pm$ 6 						&             (9, 10)			\\  
SDSS J024912.86-081525.6	&	$0.2 \times 10^{6}$				& 	53 $\pm$ 3 						&             (9, 10)			\\  
SDSS J032515.59+003408.4	&	$1.0 \times 10^{6}$				& 	50 $\pm$ 5 						&             (9, 10)			\\  
SDSS J032707.32-075639.3	&	$0.6 \times 10^{6}$				& 	75 $\pm$ 6 						&             (9, 10)			\\  
SDSS J074836.80+182154.2	&	$1.2 \times 10^{6}$				& 	42 $\pm$ 4 						&             (9, 10)			\\  
SDSS J080629.80+241955.6	&	$0.9 \times 10^{6}$				& 	71 $\pm$ 5 						&             (9, 10)			\\  
SDSS J080907.58+441641.4	&	$0.9 \times 10^{6}$				& 	65 $\pm$ 3 						&             (9, 10)			\\  
SDSS J081550.23+250640.9	&	$0.4 \times 10^{6}$				& 	65 $\pm$ 2						&             (9, 10)			\\  
SDSS J082325.91+065106.4	&	$0.5 \times 10^{6}$				& 	55 $\pm$ 7						&             (9, 10)			\\  
SDSS J082347.95+060636.2	&	$1.1 \times 10^{6}$				& 	69 $\pm$ 9						&             (9, 10)			\\  
SDSS J082443.28+295923.5	&	$0.2 \times 10^{6}$				& 	107 $\pm$ 3						&             (9, 10)			\\  
SDSS J082912.67+500652.3	&	$0.6 \times 10^{6}$				& 	60 $\pm$ 2						&             (9, 10)			\\  
SDSS J083346.04+062026.6	&	$0.3 \times 10^{6}$				& 	45 $\pm$ 4						&             (9, 10)			\\  
SDSS J091032.80+040832.4	&	$0.1 \times 10^{6}$				& 	72 $\pm$ 12						&             (9, 10)			\\  
SDSS J092700.53+084329.4	&	$1.8 \times 10^{6}$				& 	100 $\pm$ 10						&             (9, 10)			\\  
SDSS J093147.25+063503.2	&	$1.9 \times 10^{6}$				& 	52 $\pm$ 9						&             (9, 10)			\\  
SDSS J093147.25+063503.2	&	$1.7 \times 10^{6}$				& 	35 $\pm$ 6						&             (9, 10)			\\  
SDSS J093829.38+034826.6	&	$0.7 \times 10^{6}$				& 	56 $\pm$ 7						&             (9, 10)			\\  
SDSS J094057.19+032401.2	&	$0.9 \times 10^{6}$				& 	82 $\pm$ 3						&             (9, 10)			\\  
SDSS J094529.36+093610.4	&	$1.7 \times 10^{6}$				& 	76 $\pm$ 2						&             (9, 10)			\\  
SDSS J095151.82+060143.7	&	$0.4 \times 10^{6}$				& 	76 $\pm$ 6						&             (9, 10)			\\  
SDSS J101108.40+002908.7	&	$1.5 \times 10^{6}$				& 	55 $\pm$ 5						&             (9, 10)			\\  
SDSS J101627.32-000714.5	&	$0.4 \times 10^{6}$				& 	55 $\pm$ 7						&             (9, 10)			\\  
SDSS J102124.87+012720.3	&	$1.2 \times 10^{6}$				& 	78 $\pm$ 3						&             (9, 10)			\\  
SDSS J102348.44+040553.7	&	$0.2 \times 10^{6}$				& 	91 $\pm$ 13						&             (9, 10)			\\  
SDSS J103518.74+073406.2	&	$0.7 \times 10^{6}$				& 	109 $\pm$ 4						&             (9, 10)			\\  
SDSS J105755.66+482502.0	&	$0.5 \times 10^{6}$				& 	45 $\pm$ 2						&             (9, 10)			\\  
SDSS J111031.61+022043.2	&	$1.1 \times 10^{6}$				& 	77 $\pm$ 3						&             (9, 10)			\\  
SDSS J111749.17+044315.52	&	$0.6 \times 10^{6}$				& 	69 $\pm$ 5						&             (9, 10)			\\  
SDSS J112526.51+022039.0	&	$0.9 \times 10^{6}$				& 	87 $\pm$ 5						&             (9, 10)			\\  
SDSS J114339.49-024316.3	&	$2.3 \times 10^{6}$				& 	97 $\pm$ 5						&             (9, 10)			\\  
SDSS J114343.76+550019.3	&	$0.9 \times 10^{6}$				& 	31 $\pm$ 2						&             (9, 10)			\\  
SDSS J114439.34+025506.5	&	$0.3 \times 10^{6}$				& 	47 $\pm$ 4						&             (9, 10)			\\  
SDSS J114633.98+100244.9	&	$0.9 \times 10^{6}$				& 	62 $\pm$ 8						&             (9, 10)			\\  
SDSS J121518.23+014751.1	&	$1.0 \times 10^{6}$				& 	81 $\pm$ 3						&             (9, 10)			\\  
 \botrule
\end{tabular} \label{sigmas}}
\end{table}

\begin{table}[t!]
\begin{minipage}{\textwidth}
 \centering
\footnotesize
\begin{tabular}{lcccc}
\toprule
Name		 			&  $M_\mathrm{BH}$ $^{a}$		&	$\sigma$								&	References												\\
						&	[M$_{\odot}$]  					&        [km s$^{-1}$]  						&  															\\ 
\colrule
SDSS J122342.81+581446.1	&	$1.1 \times 10^{6}$				& 	45 $\pm$ 2						&             (9, 10)			\\  
SDSS J124035.81-002919.4	&	$0.9 \times 10^{6}$				& 	56 $\pm$ 3						&             (9, 10)			\\  
SDSS J131310.12+051942.1	&	$0.2 \times 10^{6}$				& 	74 $\pm$ 3						&             (9, 10)			\\  
SDSS J131310.12+051942.1	&	$0.3 \times 10^{6}$				& 	65 $\pm$ 3						&             (9, 10)			\\  
SDSS J131651.29+055646.9	&	$0.9 \times 10^{6}$				& 	82 $\pm$ 3						&             (9, 10)			\\  
SDSS J131659.37+035319.8	&	$0.5 \times 10^{6}$				& 	81 $\pm$ 7						&             (9, 10)			\\  
SDSS J131926.52+105610.9	&	$0.7 \times 10^{6}$				& 	47 $\pm$ 3						&             (9, 10)			\\  
SDSS J143450.62+033842.5	&	$0.5 \times 10^{6}$				& 	57 $\pm$ 3						&             (9, 10)			\\  
SDSS J144052.60-023506.2	&	$0.8 \times 10^{6}$				& 	73 $\pm$ 8						&             (9, 10)			\\  
SDSS J144705.46+003653.2	&	$1.9 \times 10^{6}$				& 	64 $\pm$ 4						&             (9, 10)			\\  
SDSS J150754.38+010816.7	&	$1.4 \times 10^{6}$				&   132 $\pm$ 3						&             (9, 10)			\\  
SDSS J155005.95+091035.7	&	$1.2 \times 10^{6}$				&     78 $\pm$ 6						&             (9, 10)			\\  
SDSS J161751.98-001957.4	&	$0.5 \times 10^{6}$				&     65 $\pm$ 6						&             (9, 10)			\\  
SDSS J162403.63-005410.3	&	$1.1 \times 10^{6}$				&     94 $\pm$ 4						&             (9, 10)			\\  
SDSS J162636.40+350242.0	&	$0.3 \times 10^{6}$				&     52 $\pm$ 1						&             (9, 10)			\\  
SDSS J163159.59+243740.2	&	$0.2 \times 10^{6}$				&     66 $\pm$ 2						&             (9, 10)			\\  
SDSS J170246.09+602818.9	&	$1.5 \times 10^{6}$				&     81 $\pm$ 7						&             (9, 10)			\\  
SDSS J172759.15+542147.0	&	$0.5 \times 10^{6}$				&     67 $\pm$ 8						&             (9, 10)			\\  
SDSS J205822.14-065004.3	&	$1.2 \times 10^{6}$				&     58 $\pm$ 3						&             (9, 10)			\\  
SDSS J221139.16-010535.0	&	$2.0 \times 10^{6}$				&     68 $\pm$ 7						&             (9, 10)			\\  
SDSS J230649.77+005023.4	&	$1.3 \times 10^{6}$				&     65 $\pm$ 3						&             (9, 10)			\\  
SDSS J233837.10-002810.3	&	$0.8 \times 10^{6}$				&     56 $\pm$ 2						&             (9, 10)			\\  
SDSS J234807.14-091202.6	&	$1.6 \times 10^{6}$				&     80 $\pm$ 7						&             (9, 10)			\\  
2XMM J002133.3-150751		&	$1.2 \times 10^{6}$				&	83.8 $\pm$ 3.8	$^{c}$				&		(11)								\\
2XMM J011356.4-144239		&	$0.9 \times 10^{6}$				& 	116.6 $\pm$ 0.4 $^{c}$				&		(11)								\\
2XMM J013612.5+154957	&	$1.2 \times 10^{6}$				& 	81.3 $\pm$ 1.7	$^{c}$				&		(11)								\\
2XMM J032459.9-025612		&	$0.2 \times 10^{6}$				& 	43.9 $\pm$ 4.3						&		(11)							\\
2XMM J120143.6-184857		&	$0.3 \times 10^{6}$				& 	38.3 $\pm$ 5.5	$^{c}$				&		(11)								\\
2XMM J123316.6+000512	&	$2.5 \times 10^{6}$				& 	95.7 $\pm$ 17.4 $^{c}$				&		(11)							\\
2XMM J130543.9+181355	&	$0.4 \times 10^{6}$				& 	52.3 $\pm$ 3.8	$^{c}$				&		(11)							\\
2XMM J134736.4+173404	&	$0.3 \times 10^{6}$				& 	88.1 $\pm$ 5.7						&		(11)							\\
2XMM J213152.8-425130		&	$0.2 \times 10^{6}$				& 	134.5 $\pm$ 5.5 $^{c}$				&		(11)							\\
2XMM J235509.6+060041	&	$1.2 \times 10^{6}$				& 	63.0 $\pm$ 22.1 $^{c}$				&		(11)							\\ 
 \botrule
 \end{tabular}
\justify
The uncertainty on the BH masses of POX 52 and the \cite{2007ApJ...670...92G} and \cite{2016ApJ...821...48H} sources is taken as 0.5 dex (e.g., \citealt{2014ApJ...789...17H}). \\
$^{a}$ BH mass measured using reverberation mapping. $^{b}$ Stellar velocity dispersion estimated from the gas velocity dispersion measured using the [NII] line. $^{c}$ Stellar velocity dispersion estimated from the width of the [OIII] line as $\sigma$ = FWHM$_\mathrm{[OIII]}/2.35$ (\citealt{2016ApJ...821...48H}).\\
REFERENCES. (1) \citealt{2002AJ....124.2607B}; (2) \citealt{2017ApJ...836..237N}; (3) \citealt{2003ApJ...588L..13F}; (4) \citealt{2005ApJ...632..799P}; (5) \citealt{2004ApJ...607...90B}; (6) \citealt{2015ApJ...809L..14B}; (7) \citealt{2016ApJ...829...57B}; (8) \cite{2016ApJ...831....2B}; (9) \citealt{2007ApJ...670...92G}; (10) \citealt{2011ApJ...739...28X}; (11) \citealt{2016ApJ...821...48H}.
\end{minipage}
\end{table}

\clearpage

\subsection{Seed BHs at high redshifts}	
\label{highz}
The largest samples of IMBHs have been found in the local Universe. Although a few hundred sources have been identified, it is not yet clear which seeding mechanism of SMBHs dominated in the early Universe. To better understand the formation of SMBHs at high redshifts, several campaigns have aimed at discovering faint AGN at $z \geq$ 5. Optical and IR surveys (e.g., GOODS\footnote{The Great Observatories Origins Deep Survey, \url{http://www.stsci.edu/science/goods/}}, CANDELS\footnote{Cosmic Assembly Near-Infrared Deep Extragalactic Legacy Survey, \url{http://candels.ucolick.org/index.html}}), which provide an estimate of the photometric redshift, have been combined with deep X-ray surveys (e.g., the CDF-S; \citealt{2011ApJS..195...10X}); however, no AGN candidates have convincingly been detected at $z \geq$ 5 individually nor via stacking (e.g., \citealt{2013ApJ...778..130T}; \citealt{2015MNRAS.448.3167W}; \citealt{2016MNRAS.463..348V}). This is in agreement with the extrapolation of the 3 $\leq z \leq$ 5 X-ray luminosity function to $z \geq$ 5, which predicts $<$ 1 AGN in the CDF-S (\citealt{2015MNRAS.453.1946G}). The use of NIR photometry has allowed us to reach fainter X-ray sources than direct X-ray surveys: using an \textit{H}-band luminosity selection, \cite{2015A&A...578A..83G} identified three faint AGN candidates at $z >$ 6 with X-ray luminosities in the 2-10 keV band above $10^{43}$ erg s$^{-1}$. Two of these high-$z$ AGN candidates are also selected from the CANDELS/GOODS-South survey and identified as potential direct collapse BHs by \cite{2016MNRAS.459.1432P}, who used a novel photometric method combined with radiation-hydrodynamic simulations that predict a steep IR spectrum. However, none of these candidates was identified by \cite{2016ApJ...823...95C} using different thresholds, which indicates that the identification of high-$z$ AGN candidates is very sensitive to the selection procedure (\citealt{2017MNRAS.466.2131P}). 

The dearth of detections at $z >$ 6 could be explained by a low BH occupation fraction, by smaller BHs predicted by the $M_\mathrm{BH}-M_\mathrm{*}$ relation (\citealt{2015ApJ...813...82R}; \citealt{2016ApJ...820L...6V}) than by the $M_\mathrm{BH}-M_\mathrm{bulge}$ relations, or by heavy obscuration (e.g., \citealt{2007A&A...463...79G}; \citealt{2009ApJ...693..447F}; \citealt{2009ApJ...696..110T}; \citealt{2017ApJ...837...19C}). The short phases of super-Eddington growth, followed by longer periods of quiescence, expected to occur in IMBHs (e.g., \citealt{2005ApJ...633..624V}; \citealt{2014ApJ...784L..38M}; \citealt{2016MNRAS.458.3047P}) could also decrease the probability of detecting accreting BHs (\citealt{2017MNRAS.466.2131P}). Modeling the X-ray emission of accreting BHs at z $\sim$6 and taking into account super-Eddington accretion, \cite{2017MNRAS.466.2131P} found that faint AGN progenitors at z $\sim$6 should be luminous enough to be detected in current X-ray surveys even when accounting for maximum obscuration. The authors concluded that the limited number of high-$z$ detections is caused by a low \textit{active} BH occupation fraction that results from the short episodes of super-Eddington growth and suggested that wide-area surveys with shallow sensitivities such as COSMOS Legacy (instead of deeper, smaller-area surveys as the CDF-S) are better for detecting the progenitors of SMBHs at high-$z$ (as also concluded by \citealt{2016ApJ...817...20M}; Mezcua et al. in preparation).  

\begin{figure*}
\includegraphics[width=\textwidth]{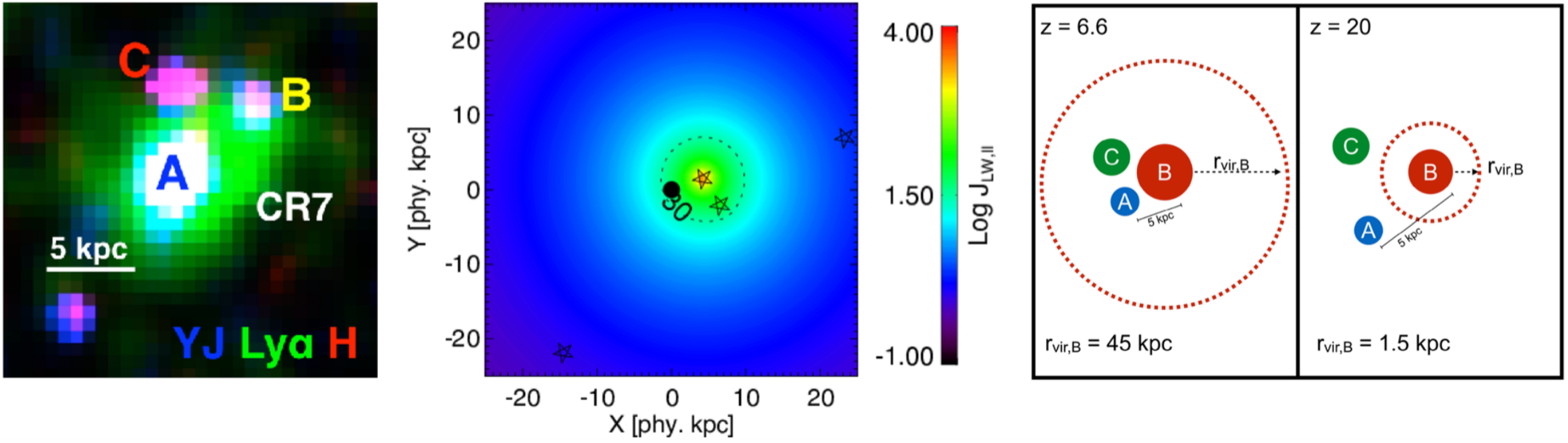}
\caption{\textbf{Left:} Observations of CR7 from \cite{2015ApJ...808..139S}. False color composite image of CR7 constructed using NB921/Suprime-cam imaging along with \textit{F110W} (YJ) and \textit{F160W} (H) filters from \textit{HST/WFC3}. The figure demonstrates the extreme blue nature of component A as compared to components B and C that are much redder. \textbf{Middle:} Simulation analogue from \cite{2014MNRAS.443..648A}. Lyman-Werner radiation contour projection in the \textit{x-y} plane for a direct collapse BH candidate (black) with a similar arrangement of neighboring galaxies (star symbols) as CR7. \textbf{Right:} evolution (not to scale). The two panels, at $z = 6.6$ (left) and $z = 20$ (right), show the evolution of the virial radius of CR7's host halo, which has a mass of $\sim10^{12}$ M$_{\odot}$ at $z = 6.6$. The stellar radiation required for direct collapse BH formation in source A is produced by source B at $z \sim$20. Source A, powered by accretion on to a direct collapse BH, later merges with the larger halo hosting source B. Figures and caption from \cite{2015ApJ...808..139S}, \cite{2016MNRAS.460.4003A}. Reproduced with permission from the RAS. \copyright AAS. Reproduced with permission.}
\label{CR7}
\end{figure*}

The detection of a strong Ly $\alpha$ emission line, virtually the only line available to confirm high redshifts, has been commonly used to detect galaxies at $z > 6$ (e.g., \citealt{2009ApJ...706.1136O,2010ApJ...723..869O}; \citealt{2013Natur.502..524F}; \citealt{2015ApJ...804L..30O}) and has recently emerged as another powerful tool to discover seed BHs at high-$z$. Those galaxies in which the formation of Pop III stars has recently taken place should emit strong Ly $\alpha$ and He II lines but no metal lines (e.g., \citealt{2015ApJ...801L..28K}). Similarly, pristine haloes, where direct collapse BHs can form in the presence of intense Lyman-Werner radiation, are expected to cool predominantly via Ly $\alpha$ line emission (e.g., \citealt{2013MNRAS.432.3438A}; \citealt{2013MNRAS.436.2989L}; \citealt{2014MNRAS.440.1263Y}). 
The discovery of strong He II line emission from the most luminous Ly $\alpha$ emitter at $z > 6$ could constitute the first detection of a (high-$z$) seed BH using this method (\citealt{2015ApJ...808..139S}). The source, CR7, is found in the COSMOS field with a redshift $z = 6.6$ (\citealt{2015MNRAS.451..400M}) and is spatially extended ($\sim$16 kpc; \citealt{2015ApJ...808..139S}). Its Ly $\alpha$ and He II lines are narrow (FWHM $\leq$ 200 km s$^{-1}$), which disfavors their origin from an AGN or Wolf-Rayet stars, and no metal lines are detected. This initially suggested that CR7 could host a population of Pop III stars. However, the finding that CR7 is formed by a blue galaxy (component A) lying close to two redder galaxies (components B and C) that could provide the Lyman-Werner radiation necessary to suppress star formation in component A seems to favor the direct collapse scenario (see Fig.~\ref{CR7}; \citealt{2015ApJ...808..139S}; \citealt{2015MNRAS.453.2465P}; \citealt{2016MNRAS.460.4003A}; \citealt{2016ApJ...823...74D}; \citealt{2016MNRAS.462.2184H}; \citealt{2016MNRAS.460.3143S}). Using deeper IR observations, \cite{2016arXiv160900727B} found evidence for metal enrichment in component A and claimed that this rules out the presence of Pop III stars or a pristine direct collapse BH. The authors suggested alternative scenarios such as the presence of a low-mass AGN or a young, low-metallicity starburst galaxy; however, using analytic models \cite{2017arXiv170200407A} showed that signatures of metals do not rule out the existence of a direct collapse BH: metal pollution of the direct collapse BH is inevitable, but the BH could form (in component A) before it is metal polluted by the same galaxies that provide the intense UV radiation required to prevent the formation of young stars (components B and C). Further observations are required to find more sources like CR7 and better constrain their nature and the conditions under which they form. 
Once the upcoming \textit{James Webb Space Telescope} (\textit{JWST}) comes online, color-color cuts in the mid-IR regime could also provide unambiguous detections of direct collapse BHs that have acquired a stellar component, termed as '\textit{obese black hole galaxies}' (OBGs). Up to 10 such OBG candidates are predicted to be detected in the CANDELS field at $z$ = 6-10 (\citealt{2013MNRAS.432.3438A}; \citealt{2017ApJ...838..117N}). This will allow us to discriminate between the two main formation mechanisms of seed BHs in the early Universe, and possibly understand the onset of the $M_\mathrm{BH}-\sigma$ relation.

\subsection{Other pathways to detection}
\label{others}
\begin{itemize}[leftmargin=0cm,itemindent=.5cm,labelwidth=\itemindent,labelsep=0cm,align=left]
\item  \textit{Tidal disruption events}\\
\label{TDEs}
Tidal disruption events (TDEs) occur when a star passing too close (within the tidal disruption radius) of a BH is ripped apart by a tidal force that exceeds the star's self-gravity (\citealt{1975Natur.254..295H}). When the bound debris is accreted by the BH, it generates a powerful flare observable from radio to $\gamma$-rays (\citealt{1990Sci...247..817R}) and that peaks in UV or soft X-rays. The flare emission declines on the timescale of months to years, producing a long-term lightcurve with a typical time decay of $t^{-5/3}$ (\citealt{2015JHEAp...7..148K}). For SMBHs above 10$^{8}$ M$_{\odot}$, the tidal disruption radius is smaller than the Schwarschild radius and solar-mass stars are not disrupted but swallowed whole without emitting any flares (\citealt{1988Natur.333..523R}; \citealt{1990ApJ...351...38C}; \citealt{2011Sci...333..203B}; \citealt{2012PhRvD..85b4037K}). SMBHs with M$_\mathrm{BH} < 10^{8}$ M$_{\odot}$ can disrupt solar-type stars; however, the disruption of compact stars such as a white dwarf can be only produced by BHs  $< 10^{5}$ M$_{\odot}$ (\citealt{2011ApJ...743..134K}). IMBHs, with BH masses below 10$^{6}$ M$_{\odot}$, should have higher rates of stellar disruption than SMBHs (\citealt{2004ApJ...600..149W}; \citealt{2016MNRAS.455..859S}); nonetheless, most TDEs ($\sim$50; \citealt{2015JHEAp...7..148K}) have been associated with SMBHs (e.g., \citealt{1999A&A...343..775K}; \citealt{2004ApJ...604..572H}; \citealt{2011ApJ...738...52L,2015ApJ...811...43L}; \citealt{2011ApJ...741...73V}; \citealt{2012ApJ...753...77C}; \citealt{2012Natur.485..217G}; \citealt{2012A&A...541A.106S,2017A&A...598A..29S}; \citealt{2014ApJ...780...44C}; \citealt{2014MNRAS.445.3263H}; \citealt{2015ApJ...798...12V}; see \citealt{2015JHEAp...7..148K} for a review). 

The few TDEs suggested to occur in IMBHs come from events observed in dwarf galaxies: the gamma-ray burst GRB 110328A, or \textit{Swift} J164449.3+573451, discovered by the \textit{Swift} Burst Alert Telescope (BAT) took place in a dwarf galaxy at $z$ = 0.354 and with log M$_\mathrm{*} = 9.14$ (\citealt{2011Sci...333..199L}; \citealt{2015ApJ...808...96Y}). A BH of mass $10^{5}-10^{7}$ M$_{\odot}$ is estimated to be the responsible for this TDE (\citealt{2011Natur.476..421B}; \citealt{2011ApJ...743..134K}; \citealt{2011Sci...333..199L}; \citealt{2011ApJ...738L..13M}; \citealt{2012A&A...548A...3A}; \citealt{2015ApJ...808...96Y}). An X-ray flare detected in the dwarf galaxy WINGS J1348 (M$_{*}\sim3 \times 10^{8}$ M$_{\odot}$; \citealt{2014MNRAS.444..866M}), in the Abell cluster 1795, indicates the tidal disruption of a star by an IMBH of log M$_\mathrm{BH} \sim 5.3-5.7$ M$_{\odot}$ (\citealt{2013MNRAS.435.1904M,2014MNRAS.444..866M}; \citealt{2014ApJ...781...59D}; see Fig.~\ref{TDE}). \cite{2013ApJ...769...85S} suggested that the tidal disruption of a white dwarf by an IMBH of $\sim10^{4}$ M$_{\odot}$ is the responsible for the gamma-ray burst GRB060218 and associated supernova SN2006aj observed in a dwarf galaxy at a redshift $z=0.0335$ (e.g., \citealt{2006Natur.442.1014S}; \citealt{2007MNRAS.375L..36G}; \citealt{2007ApJ...658L...5M}). The variability of RBS 1032, a \textit{ROSAT} X-ray source associated with the inactive dwarf galaxy SDSS J114726.69+494257.8, is suggested to be an accretion flare from a possible IMBH (\citealt{2014ApJ...792L..29M}).
TDEs in dwarf galaxies may thus be a potential tool for detecting IMBHs, even during quiescence, and for constraining their occupation fraction out to high redshifts (\citealt{2016arXiv161101386F}).\\

\begin{figure*}
\includegraphics[width=\textwidth]{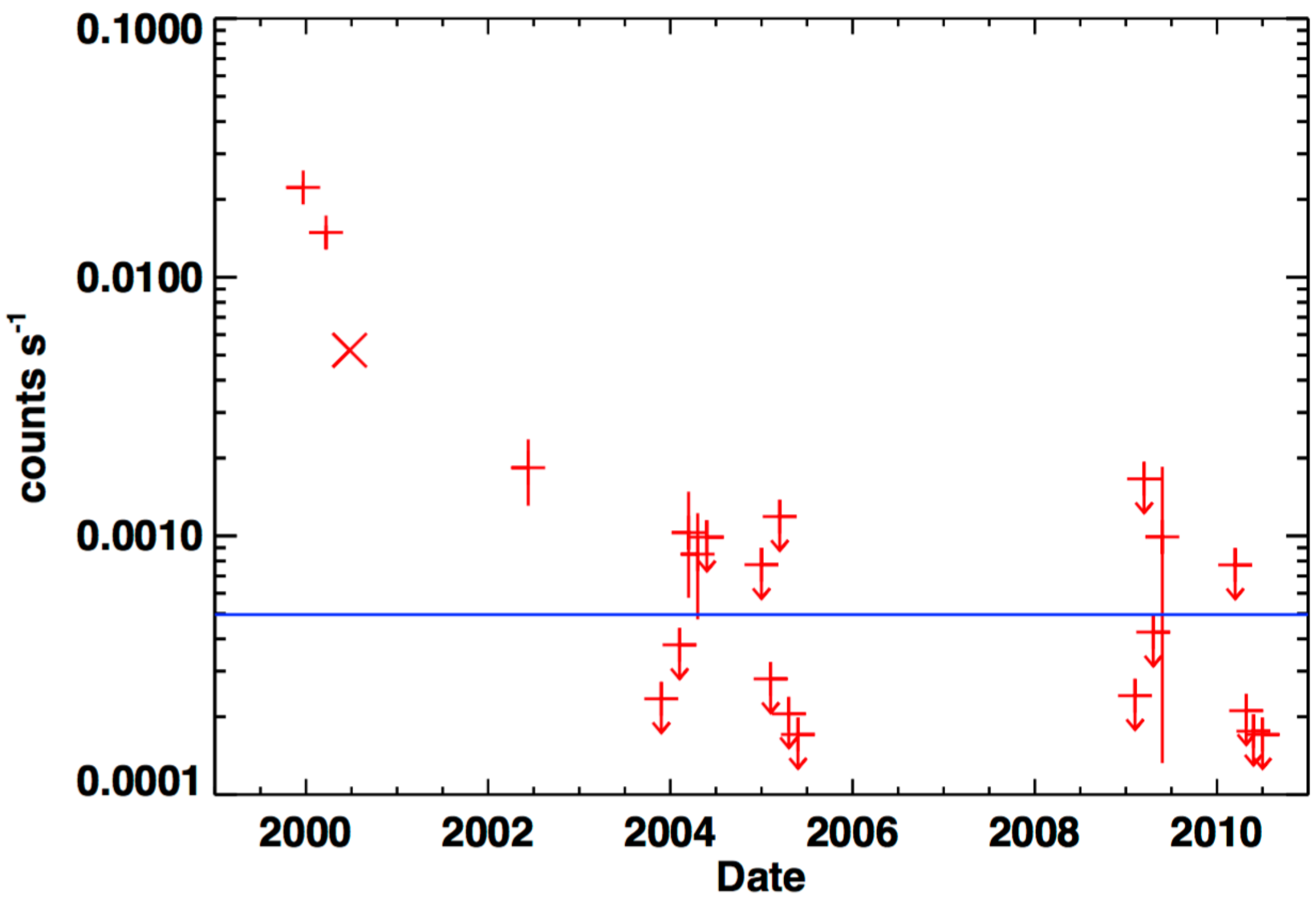}
\caption{\textit{Chandra} soft (0.3-2 keV) count rate evolution for WINGS J1348. Count rate uncertainty is indicated by cross extent. Arrows indicate 2$\sigma$ upper limits. After 2004, individual observations within a given year are offset along the X-axis to avoid overlap. The blue horizontal line is the median 2$\sigma$ upper limit in the hard (2-8 keV) band. Figure and caption from \cite{2013MNRAS.435.1904M}. Reproduced with permission from the RAS.}
\label{TDE}
\end{figure*}

\item \textit{Gravitational waves}\\
\label{gravitationalwaves}
The double detection by the advanced LIGO\footnote{Laser Interferometer Gravitational-wave Observatory.} of gravitational waves (GWs) produced by the coalescence of two heavy (compared to those in our Galaxy; i.e., $\geq$ 25 M$_{\odot}$) stellar-mass BHs (signal GW150914; \citealt{2016PhRvL.116f1102A}) and of two $\sim$10 M$_{\odot}$ BHs (signal GW151226; \citealt{2016PhRvL.116x1103A}) opened a new era in astronomy, revolutionizing the way in which we can detect BHs. Since the more massive the BHs the lower the frequency of the gravitational waves emitted by the binary system, the inspiral of an IMBH into a SMBH or of two IMBHs with BH masses $\geq10^{3}$ M$_{\odot}$ will produce GWs with frequencies too low for the current ground-based GW interferometers but in the observational range of future space-based interferometers (10$^{-5}$--10 Hz; e.g., \citealt{2006ApJ...646L.135F}). The ring-down GW of IMBH binaries with total masses in between $\sim200-2 \times 10^{3}$ M$_{\odot}$ would produce a significant signal-to-noise ratio in the advanced LIGO and VIRGO interferometers and the future Einstein Telescope (e.g., \citealt{2006ApJ...646L.135F}; \citealt{2010ApJ...722.1197A}; \citealt{2011GReGr..43..485G}; \citealt{2017ApJ...835..276S}) with a detection rate in the advanced detectors of $\sim$10 mergers a year. The binary formed by an IMBH and a stellar-mass BH would also be detectable by these ground-based interferometers, with an estimated detection rate of up to tens of events per year (\citealt{2016MNRAS.457.4499H}). In this case, it would be possible to claim the detection of GWs from an IMBH at 95\% confidence if the merging IMBH has a mass of at least 130 M$_{\odot}$ (\citealt{2016MNRAS.457.4499H}). The detection of IMBHs through GW emission seems to be thus on the horizon.\\


\item \textit{Accretion disks in AGN}\\
IMBHs could also be found in the accretion disks of AGN if they grow in a manner similar to planets in protoplanetary disks (\citealt{2011MNRAS.417L.103M}, \citeyear{2012MNRAS.425..460M}, \citeyear{2014MNRAS.441..900M}; \citealt{2016ApJ...819L..17B}). The accretion disks around SMBHs contain nuclear cluster objects (i.e., stars, compact objects, XRBs) that can collide with each other, merge, and exchange angular momentum with the gas in the disk. This can make them migrate within the disk and accrete from it, which may form an IMBH seed (\citealt{2011MNRAS.417L.103M}; \citealt{2016ApJ...819L..17B}). The GWs emitted by the rate of stellar-mass BH mergers required to produce the IMBH could naturally account for the heavy stellar-mass BHs observed by LIGO (e.g., \citealt{2017MNRAS.464..946S}; \citealt{2017arXiv170207818M}) and is also consistent with the inferred 9-240 Gpc$^{-3}$ yr$^{-1}$ rate from LIGO (\citealt{2016PhRvX...6d1015A}).

IMBH seeds in AGN disks can then grow very rapidly and reach super-Eddington accretion rates via slow collisions with nuclear cluster objects in the disk (\citealt{2012MNRAS.425..460M}). If the IMBH exhausts the nuclear cluster objects around it, it can migrate in the disk and expand its feeding zone so that super-Eddington growth can continue. This growth mechanism of IMBHs in AGN disks can open a gap in the disk, which would result in several observational signatures such as oscillations on the broad Fe K$\alpha$ line profile of the AGN caused by the presence of the secondary BH (i.e., the IMBH; \citealt{2013MNRAS.432.1468M}) or a reduction of the ionizing continuum luminosity of the AGN if the disk is removed to large radii. The later can yield luminosities consistent with those of LINERs and change the optical lines ratio from AGN-like to LINER-like (\citealt{2011MNRAS.413L..24M}), suggesting that many LINERs could consist of binary BHs: a SMBH and a secondary BH (the IMBH) that opens a large cavity in the SMBH disk. Such binary systems would emit GWs during the IMBH inspiral, which could be detected with the up-coming eLISA (\citealt{2014MNRAS.441..900M}). The non-detection of IMBH-SMBH binaries by eLISA would imply that this is not an efficient channel for producing IMBHs.

The blackbody spectrum of the accretion disk of an IMBH peaks in the soft X-ray band ($\sim$0.1-1 keV for BH masses ranging 10$^{4}-10^{2}$ M$_{\odot}$). IMBHs in AGN disks could thus also be detected as an excess of soft X-rays relative to the soft X-ray emission expected from an extrapolation of a power-law fit from hard X-rays (\citealt{2014MNRAS.441..900M}). Soft X-rays could also be generated by the bow shock associated with the headwind of a non-gap-opening IMBH on a retrograde orbit (\citealt{2014MNRAS.441..900M}). Additionally, IMBHs in AGN disks could also produce QPOs, which have been occasionally observed in AGN (\citealt{2010A&A...524A..26C}), asymmetric X-ray intensity distributions that may be detected as AGN transits (e.g., \citealt{1998ApJ...501L..29M}; \citealt{2010A&A...517A..47M}; \citealt{2011ApJ...742L..29R}), and UV/X-ray flares caused by a TDE.\\

\item \textit{High velocity clouds}\\
The presence of an IMBH has been also proposed in the dense interstellar gas surrounding the nucleus of the Milky Way within a few hundred parsecs (\citealt{2016ApJ...816L...7O}). Using the Nobeyama Radio Observatory 45 m radio telescope, \cite{2016ApJ...816L...7O} found a compact ($<$5 pc) molecular cloud named CO-040-022 whose kinematical structure consists of an intense region with a shallow velocity gradient plus a less intense high-velocity wing. The kinematical appearance and very high velocity width ($\sim$100 km s$^{-1}$) of CO-040-022 can be explained by a gravitational kick imparted by a compact object not visible at other wavelengths and with a mass of 10$^{5}$ M$_{\odot}$ (\citealt{2016ApJ...816L...7O}). This suggests the presence of an IMBH and opens a new window in the search for these objects: whether they are active or not, they leave a kinematic imprint when encountering a molecular cloud. The finding of compact high-velocity features when studying molecular line kinematics could thus yield the detection of many more IMBHs. 

\end{itemize}

\begin{figure*}
\includegraphics[width=\textwidth]{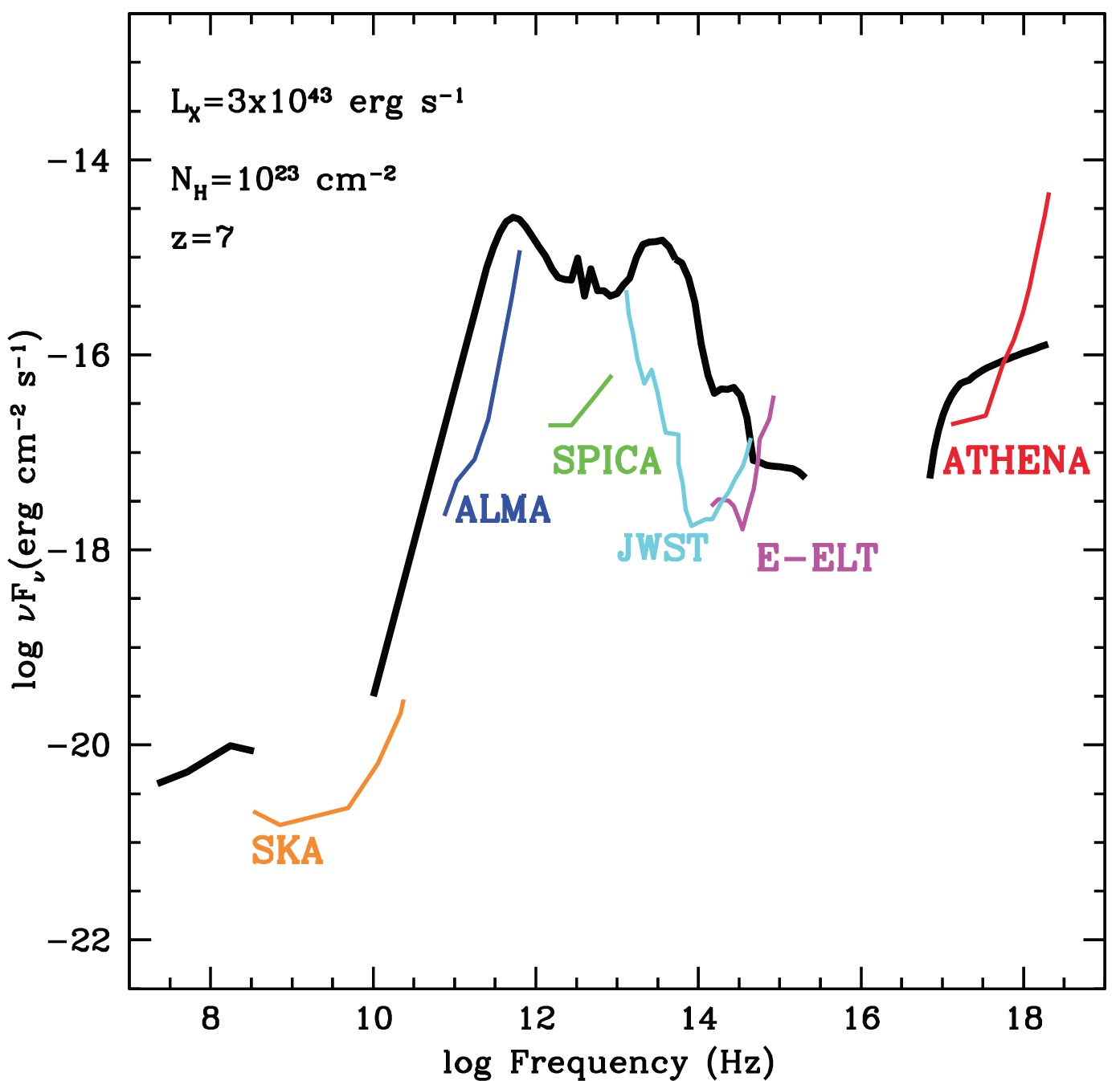}
\caption{Broad-band SED of a moderate-luminosity obscured AGN at $z=7$ that will be observable in the \textit{ATHENA} surveys. The thick black line is that of an obscured AGN with similar luminosity and obscuring column density in the COSMOS survey (\citealt{2011A&A...534A.110L}) redshifted to $z=7$. The 3$\sigma$ sensitivities (for a typical survey exposure) of SKA, ALMA, SPICA, \textit{JWST} and E-ELT are also shown. Figure and caption from \cite{2013arXiv1306.2325A} and \cite{2016PASA...33...54R}. Reproduced with permission.}
\label{SED}
\end{figure*}

\section{Conclusion and Outlook}	
According to cosmological models of BH growth, the determination of the number of present-day IMBHs can elucidate how seed BHs formed in the early Universe: either from the death of Pop III stars, the direct collapse of pristine gas, or by stellar mergers in dense stellar clusters (e.g., \citealt{2010A&ARv..18..279V}). A few hundreds of IMBH candidates have been now identified in a variety of objects, from dwarf galaxies and ULXs to TDEs and high velocity clouds; nonetheless the BH occupation fraction has not yet been constrained to a level that allows us to draw conclusive results about the dominant seeding mechanism at high redshift. \cite{2013ApJ...775..116R} provide an AGN fraction for optically selected dwarf galaxies at $z < 0.055$ of 0.5\%, a value which is not corrected for incompleteness and which is much lower than the lower limit of $> 20$\% found by \cite{2015ApJ...799...98M} for dwarf early-type galaxies with Eddington ratios down to 10$^{-4}$. \cite{2016ApJ...831..203P} apply an incompleteness correction to their sample of dwarf galaxies with $10^{9}$ M$_{\odot}$ $< M_\mathrm{*} < 3 \times 10^{9}$ M$_{\odot}$ at $0.1 < z < 0.6$ and find an active fraction of 0.6\%-3\%, which is in agreement with the 3.1\% obtained from semi-analytic models of galaxy formation that seed halos with 10$^{4}$ M$_{\odot}$ BHs. However, to obtain the true BH occupation fraction we need a larger cosmological volume, a wider range of masses, and to know the distribution of Eddington ratios across the mass scale. The distribution of AGN X-ray detections can be used to infer the BH occupation fraction, e.g., using a mock catalog with 15,000 galaxies and $\sim$300 X-ray AGN, \cite{2015ApJ...799...98M} find an X-ray AGN fraction of 11.9\% in dwarf galaxies with full occupation but of 6.1\% if they have half occupation. Wide-area X-ray surveys, such as the one that will be carried out by the Wide Field Imager instrument onboard of the up-coming \textit{ATHENA} X-ray satellite, can provide the larger volumes required to constrain the BH occupation fraction.

The BH masses of the IMBH candidates so far found are always in the range $10^{4}-10^{6}$ M$_{\odot}$, which is the typical mass of the seed BHs formed from direct collapse and adopted in large-scale cosmological simulations that show how their growth could explain the highest-redshift SMBHs (e.g., \citealt{2011ApJ...742...13B}; \citealt{2011ApJ...738...54K}; \citealt{2012MNRAS.420.2662D}). This, together with the discovery of CR7 (e.g., \citealt{2015ApJ...808..139S}; \citealt{2016MNRAS.459.1432P}) and the plume-like distribution of IMBHs/low-mass AGN around $M_\mathrm{BH} \sim10^{5}-10^{6}$ M$_{\odot}$ on the $M_\mathrm{BH}-\sigma$ relation (see Sect.~\ref{dwarf}, Fig.~\ref{Msigma}), seems to favor the direct collapse scenario (e.g., \citealt{2012NatCo...3E1304G}). However, the number density of BHs expected from direct collapse is relatively low (e.g., \citealt{2016MNRAS.463..529H}); so there might just be a bias towards the direct collapse scenario caused by the easier detection of heavy seeds with large BH masses. Light seeds formed from Pop III stars have not grown much since their formation, hence, although they should have a higher occupation fraction, their detection is harder (e.g., \citealt{2010MNRAS.408.1139V}; \citealt{2010A&ARv..18..279V}).

Additional constraints on the mass and growth of seed BHs can be provided by the X-ray non-detections of faint AGN at $z \geq$ 5 (e.g., \citealt{2013ApJ...778..130T}; \citealt{2015MNRAS.448.3167W}; \citealt{2016MNRAS.463..348V}) nor of Lyman Break Galaxies (e.g., \citealt{2011ApJ...742L...8W}; \citealt{2012ApJ...748...50C}; \citealt{2012A&A...537A..16F}). The current non-detections can be explained by a low BH occupation fraction, BH masses $\leq 10^{5}$ M$_{\odot}$ (\citealt{2016ApJ...820L...6V}), heavy obscuration, or intrinsic X-ray weakness (\citealt{2014ApJ...794...70L}). This later possibility has been also proposed to explain the unusual behavior of local low-mass AGN whose ratio of UV to X-ray is lower than 'normal' AGN and are relatively X-ray weak compared to the AGN power expected from ionized gas emission, and it would imply that these sources are governed by a different physical regime than that producing the characteristic strong X-ray-emitting corona of AGN (e.g., \citealt{2012ApJ...761...73D}; \citealt{2016ApJ...825..139P}). 

The advent of the next generation of ground and space observatories (\textit{ATHENA}, SKA, \textit{JWST}, E-ELT) and dedicated surveys will provide a giant leap on the detection of faint AGN at high redshifts and of direct collapse BHs (e.g., \citealt{2015MNRAS.453.2465P}; \citealt{2017ApJ...838..117N}); yet, BHs fainter than $3 \times 10^{43}$ erg s$^{-1}$ will be extremely difficult to detect at $z=7$ even by this new generation of telescopes (see Fig.~\ref{SED}). It is the multi-wavelength study of IMBHs in dwarf galaxies (e.g., such as the multiple-method approach performed by \citealt{2017A&A...601A..20K}), in ULXs or in globular clusters, and of AGN in high-redshift quasars the one that coupled with simulations of BH formation models will allow us to understand how seed BHs formed and grew to become the SMBHs we know today. 


\section*{Acknowledgments}
I would like to thank the anonymous referee for carefully reading the manuscript as well as Pau Diaz Gallifa and Marios Karouzos for constructive comments and illuminating discussions. I am also thankful to Bhaskar Agarwal, Vivienne Baldassare, Barry McKernan, Amy Reines, and Marta Volonteri for useful suggestions that have helped improve the manuscript and, together with James Aird, Andrea Comastri, Jenny Greene, Richard Plotkin, and David Sobral, for allowing me to reproduce their figures in this review.


\small



\bibliographystyle{apj} 
\bibliography{/Users/mmezcua/Documents/referencesALL}

\end{document}